\documentclass[preprint,11pt]{aastex}
\newcommand{\cf}{cf.,}
\newcommand{\ie}{i.e.,}

\newcommand{\etal}{et al.}
\newcommand{\muas}{\ensuremath{\mu\mbox{as}}}

\newcommand{\muasyr}{\ensuremath{\mu\mbox{as\,yr}^{-1}}}

\newcommand{\Msun}{\ensuremath{\mbox{$M_{\sun}$}}}
\newcommand{\Rsun}{\ensuremath{\mbox{$R_{\sun}$}}}
\begin{document}
\title{Selection of the SIM Astrometric Grid}
\author{D.~M.~Peterson\altaffilmark{1}}
\affil{Department of Physics and Astronomy,
SUNY, Stony Brook, NY 11794-3800}
\altaffiltext{1}{on leave at: Remote Sensing Division, 
Naval Research Laboratory,
4555 Overlook Avenue SW, Washington, DC 20375}
\email{dpeterson@astro.sunysb.edu}
\author{Y.~Liu}
\affil{Department of Mathematics,
SUNY, Stony Brook, NY 11794-3651}
\email{yliu@astro.sunysb.edu}
\and
\author{S.~Portegies~Zwart}
\affil{Astronomical Institute 'Anton Pannekoek' and 
Dept.\ of Computer Science, Univeristy of Amsterdam,
Kruislaan 403, 1098SJ Amsterdam,
the Netherlands}
\email{spz@science.uva.nl}

\begin{abstract}
We investigate the choice of stellar population for use as the
Astrometric Grid for the Space Interferometry Mission (SIM).  SIM
depends on the astrometric stability of about 2000 stars, the so called
Grid, against which the science measures are referenced.  Low
metallicity, and thus relatively high luminosity K giants are shown to
be the population of choice, when available.  The alternative, nearby G
dwarfs, are shown to be suseptable to unmodeled motions induced by
gas-giant planetary companions, should there be a significant population
of such companions..

Radial velocity filtering is quite efficient in selecting Grid members
from the K giants with yields exceeding 50\% if filtering at
30\,ms$^{-1}$ ($1\sigma$) is available.  However if the binary fraction
of the G dwarfs approaches 100\% as some studies suggest, the yield of
stable systems would be in the range of 15\% at best (with 10\,ms$^{-1}$
filtering).  Use of the initial SIM measurement as a final filter is
shown not to be critical in either case, although it could improve the
yield of stable grid members.

For a Grid composed of weak-lined K giants, the residual contamination
by large unmodeled motions will amount to about 3\% (and rises to about
6\% if a 60\,m\,s$^{-1}$ radial velocity criterion is used).  The
selective introduction of quadratic terms in the proper motion solutions
during the post-mission phase of data reduction can reduce contamination
to a remarkable 1\% or better in either case.

Analytic estimates based on circular orbits are developed which show how
these results come about.
\end{abstract}

\vfill\eject
\section{INTRODUCTION}
The Space Interferometry Mission \citep[SIM, \cf][]{BUS97}, scheduled
for launch in 2009, is designed to move the practice of Global
Astrometry to an extraordinary level of precision, with positions and
parallaxes accurate to 4 \muas\ and proper motions accurate to 2
\muasyr\ for targets as faint as $\mbox{R}=20$\,magnitudes.  Over one or
two degree fields (the ``Narrow Angle'' mode), relative precisions of a
few 100 nanoarcsec may be obtained by mission end.

To reach these levels a group of about 2000 stars, known as the
``Grid'', which is thinly but fairly uniformly spread over the sky, is
intensively observed throughout the nominal 5 year mission.  Absolute
astrometric measurements of science targets are referenced to Grid
members, while in Narrow Angle mode, orientation and scale are provided
by Grid observations.  The fundamental requirement on a Grid member is
that over the mission it exhibit no motions other than those described
as proper motion and parallax to levels significantly below the
precision levels listed above (there may be some easing of these
constraints, as discussed below).  Second only to the health of the
instrument itself, the success of the mission depends on identifying an
acceptable Grid membership and establishing its astrometric properties.

We consider here the issue of identifying the group of stars most likely
to yield an astrometrically stable Grid.  Two specific groups of stars
have been proposed: nearby G dwarfs and kiloparsec distant, weak line K
giants.  The first group, the nearby G dwarfs, would leverage on the
already extensive radial velocity screening they have undergone
\citep[\ie\ ][and references therein]{Udry98}.  The second group, the K
giants was suggested by Majewski \citep[1998 private
communication]{Patterson99}.  That these objects are at distances in
excess of 1 kpc reduces the astrometric signature when companions are
present, substantially helping the filtering problem.

We first consider the parameter space that must be probed in order to
filter these two populations, using the approximation of the circular
orbit.  With elementary analytic analysis we show not only worst case
requirements but the effect of random inclinations of the orbital planes
and random phase in the orbits (critical for periods longer than the
mission life).

The effects of finite eccentricities are not so amenable to simple
treatment and we resort to extensive Monte Carlo simulations to show
that within the expected range of eccentricity distributions, the
results are essentially as we would expect on the basis of the circular
approximation.

We also consider the possibility that SIM itself can act as one last
filter during the first set of observations of each object.
\citet[hereafter LP]{LP00}, have developed a technique that looks for
characteristic variations in the measured phase delays as a function of
wavelength using $\chi^2$ and Periodogram statistics.  SIM would be
surprisingly powerful in detecting companions to G dwarf primaries.
However, it covers no unique parameter space compared to radial velocity
searches and K-band imaging, and would be useful only to catch systems
that slipped through those ground based efforts.  For the K giant
primaries the technique can make little contribution except again as a
second check on the radial velocity coverage.

The statistical distributions describing K giant binaries, which must
evolve from main sequence systems, are considered.  We argue that about
15\% of the K giants will have evolved from systems where the current
primary was the original secondary, and where the original primary is
now a white dwarf which has undergone some mass loss.  Otherwise the
main differences in the statistical properties of these systems has to
do with the increase in radius of the primary, which we allow for
explicitly.

While we explore the effects of differing assumptions in the various
distributions, the primary conclusion, which is remarkably robust, is
that while it is possible to develop a G dwarf sample with a relatively
small residual contamination from astrometrically unstable systems, it
comes at the cost of a very low overall yield in acceptable Grid members
(assuming 100\% binaries in the original population).  On the other
hand, $R \sim 12$ weak line gK systems should provide about a 50\%
overall yield with a residual 5\% contamination of unstable Grid
members, even if the underlying population is 100\% binaries.

\cite{Gould} has recently considered the Grid problem and specifically the
use of Pop\,II K giants, using analytic developments.  Our approach is
quite complementary to his and the results reassuringly similiar on the
aspects considered in common.

\vfill\eject
\section{ANALYTIC ESTIMATES: CIRCULAR ORBITS}

Circular orbits have been used widely to characterize the problem of
astrometric detection of companions \citep[\cf][]{Booklet}.  In the current
context the approximation is very useful and we develop the various
expressions that define best- and worst-case scenarios over the relevant
range of parameters.

In both populations, main sequence G stars (``dG'') and weak line (thick
disk, possibly halo) K giants (``gKw''), we assume a 1\,\Msun\ primary.  We
confine ourselves to binary systems.  That is, we assume the potential
targets all have companions, but no more than one.  We discuss the
implications of this latter choice when we discuss the more realistic
simulations, below.

The fundamental relations, in Keplerian units, are then:
\begin{eqnarray}
a^3 = P^2 m_p (1 + q)\\
v_c = \frac{a_p}{P} = \frac{m_s}{[a m_p (1 + q)]^{1/2}}
\end{eqnarray}
where $a$ is the system semimajor axis and $a_p$ that of the primary
(both in AU's), $m_p$ is the mass of the primary ($m_p = 1$), $m_s$ that
of the secondary (both in solar masses), $q = m_s/m_p$ is the mass
ratio and $v_c$ is the circular velocity (in units of the Earth's mean
velocity, which conveniently absorbs the factor of $2\pi$).

In all cases we assume the primary light dominates the system and it is the
radial velocity and astrometric variations of the primary only that can be
detected.  This is obviously not correct when the mass ratio approaches
unity in the G dwarf case.  Since the issue is to identify those systems
that will or will not be detectable by the various techniques considered,
this approximation does not introduce significant error.  That is to say,
in the extreme case of essentially identical G dwarfs, the combination of
mass ratio, inclination and period that would render spectral lines to be
so completely blended as to escape detection as a radial velocity variable,
and yet produce significant unmodelled astrometric motions, would result in
additional contamination well below the 1\% limit on the accuracy of the
results reported here.

Again for the dG systems, we consider only those which are in their initial
main sequence phases.  In particular, we assume that the more massive
component is the brighter.

\subsection{Radial Velocity Detection}

In considering the sensitivity of radial velocity measurements to orbital
motion we develop the edge-on case, noting that since only 1-dimensional
motions are involved, finite inclination of the orbital plane always adds a
simple $\sin i$ factor ($i$ is defined to be $\pm 90^{\circ}$ for edge-on
systems, pro- or retrograde).

When the binary period is longer than the measurement interval, the issue
of just what range of orbital phases are sampled becomes important.  Then,
location in orbit becomes another random variable that must be convolved
with the distribution of inclinations, for example, in evaluating the
actual detection frequency.  To appreciate the range of this effect on
detection, resulting from randomness in orbital phase, we consider
``best'', ``worst'' and 90\% detection cases.

In this and what follows we do not worry about the distinction between a
semi-amplitude and a true rms amplitude as would be obtained from a least
squares fit to a well sampled velocity curve.

\subsubsection{Best Case}

For the edge-on case we set the total velocity range that would be sampled
during the pre-mission preparatory phase, $\Delta t_p$, equal to twice the
precision we expect to obtain for the given population, $\varepsilon_v$.  The
velocity sinusoid gives the largest variation for observations centered
on nodal passage, \ie
\begin{equation}
2\varepsilon_v = \left \{ \begin{array}{r@{,\quad}l}
		     2v_c & P < 2 \Delta t_p\\
		     2v_c\sin\frac{\pi\Delta t_p}{P} & P \geq 2 \Delta t_p.
	      \end{array}\right .
\end{equation}
In this case the phasing is optimal for detection.  For long periods the 
latter becomes 
\begin{equation}
\varepsilon_v \sim v_c \frac{\pi\Delta t_p}{P} = \pi\Delta t_p \frac{m_s}{a^2}
\end{equation}
to first order.

\subsubsection{Worst Case}

Alternatively, the smallest velocity range in a given observing period is
obtained when the system is at quadrature, \ie
\begin{equation}
2\varepsilon_v = \left \{ \begin{array}{r@{,\quad}l}
		     2v_c & P < \Delta t_p\\
		     v_c[1-\cos\frac{\pi\Delta t_p}{P}] & P \geq \Delta t_p.
	      \end{array}\right . 
\end{equation}
In the long period limit this becomes
\begin{equation}
\varepsilon_v \sim v_c \left (\frac{\pi\Delta t_p}{2P}\right )^2 
= \left (\frac{\pi\Delta t_p}{2}\right )^2 \frac{m_s}{a^{7/2}} 
[m_p(1+q)]^{1/2}.
\end{equation}

\subsubsection{90\% range}

As can be seen above, when the system period is significantly longer
than the interval during which it was scrutinized for velocity
variations, the ``best'' case phasing for detection diverges from the
``worst'' by a factor of $\pi\,\Delta t_p/4P \sim 0.8\Delta t_p/P$.  For
separations of 100\,AU this factor approaches 250.  Orbital phase
clearly plays a significant role in determining whether the system will
be detected by a radial velocity search.

\begin{figure}
\epsscale{.70}
\plotone{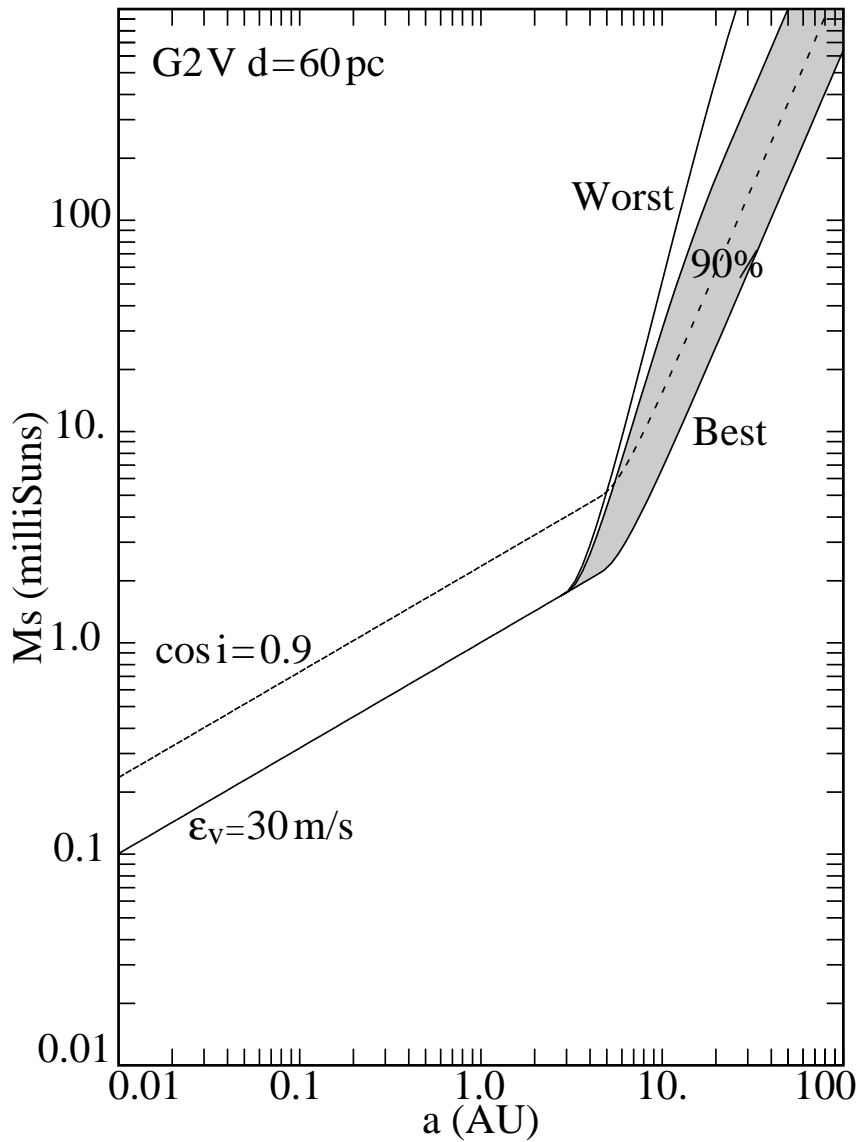}
\caption{\label{RadialVel} Radial velocity detection of
companions as a function of secondary mass and total separation.  The case
of G dwarfs at 60\,pc is taken for illustration with a velocity sensitivity
($3\sigma$) of 30\,m\,s$^{-1}$ assumed.  For periods longer than $\Delta
t_p = 5$ years we show the effects of phase in orbit including nodal
passage (``Best''), quadrature passage (``Worst'') and, by shading, the
range for 90\% detection assuming a uniform distribution of phases.  These
are for edge-on systems.  The dashed line indicates the range for detection
at the 90\% level assuming a random orientation of orbital planes.  (Beyond
5 years, this is for the ``Best'' case only).  In all cases, systems with
parameters to the upper left are predicted to be detectable.}
\end{figure}

However, the problem is to detect a certain fraction of systems, not
all, and it is elementary to calculate the radial velocity limits which
will detect, say, 90\% of the systems.  An example of this calculation
is shown in Fig.\ \ref{RadialVel} where we use the parameters we adopt
for the case of a nearby dG population.  ``Best'' and ``Worst'' cases in
the sense used here are shown by solid lines.  The region that would
encompass 90\% of the systems, assuming they are uniformly distributed
in phase, is shaded.

\subsubsection{Inclination}

All this assumes the orbit is edge-on, \ie\ $i = \pm90^{\circ}$.
Inclination effects are always present.  However, if orbital angular
momenta vectors are distributed uniformly over the sky, probably a good
assumption, then inclinations are distributed as $f(i)=\sin i$ and the
chances of seeing systems face on are very low.  In Fig.\
\ref{RadialVel} we show with a dashed line the locus that would allow
detection of 90\% of the systems assuming the inclinations are
distributed as above.  Interestingly, for long period systems, orbital
phase effects are much more important, at least at the 90\% detection
level.

\subsection{Astrometric ``Noise''}

Next, we need to define exactly what will produce an erroneous
positional measurement in the context of the Grid.  For the ideal Grid
member, specifying the position, proper motion and parallax will define
its location at any epoch.  To the extent that binary motions change
positions in ways that cannot be modeled so simply, duplicity introduces
errors into the Grid.

When these unmodeled motions exceed a certain maximum, the object will
probably be rejected as a Grid member.  This decision will in most
likelihood be made at the end of the mission, not allowing any
remediation (but see below for one proposal to soften this
circumstance).  Since a certain minimum number of acceptable Grid
members (of order 4) must be present in any 15$^{\circ}$ ``Field of
Regard'' (FoR), redundancy must be provided in the original Grid
membership.  To prevent {\it any} FoR from dropping below the critical
number of usable Grid members, the multiplier to provide the redundancy
can be fairly large \citep[\cf][]{Wade00} and correspondingly costly in
terms of observing time.

Since astrometric measurements are fundamentally two dimensional,
unmodeled motions can be present in either coordinate or shared between
them.  For our purposes we consider the largest component of any error,
regardless of how it is projected in the observational coordinate frame.

Here and in the simulations below, we remove only linear terms from the
binary motions.  In particular, we do not model the parallax term.  This
will produce a significant modeling error only for systems with periods
close to one year, since the Grid will be observed much more frequently
than twice a year and over at least 5 years.  As a result, the aliasing
will be confined to a small region in parameter space.  And, since the
effect of adding a parallax term would be to absorb binary system
motions that cannot be modeled by the linear terms provided (at the
price of reduced parallax accuracy), we will tend to report more
astrometric instability than would be observed in practice.  Again, we
expect the effect to be very small.

\subsubsection{Face-On Systems}

\begin{figure}
\plotone{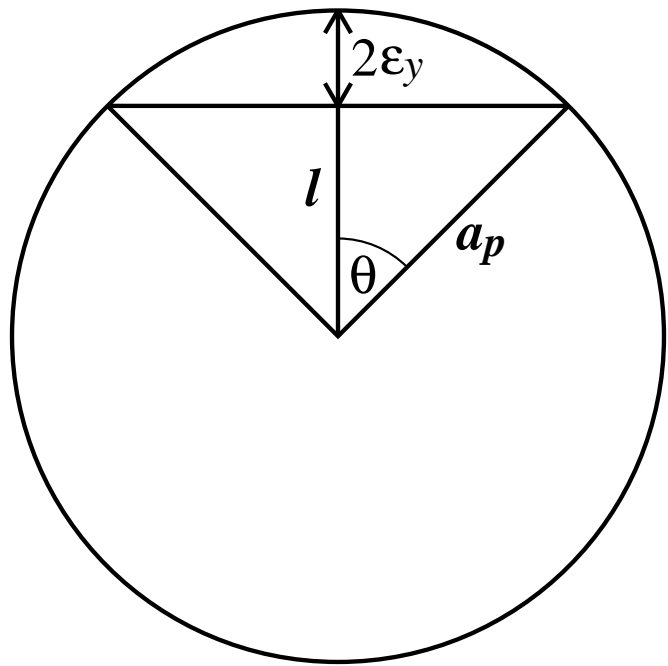}
\caption{\label{Circle1} The astrometric ``Worst'' case.  The motion along
the $y$ axis has no linear component to remove, we equate the amplitude to
twice the maximum allowable noise, $\varepsilon_y$.}
\end{figure}

For circular systems the ``Curvature Noise'' (unmodeled astrometric
motions) analysis follows \citet{Hajian99} and is illustrated in Fig.\ 
\ref{Circle1} for the face-on case.  Here, orbital motion causes
displacements in both $x$ and $y$ coordinates.  Over a finite time
interval corresponding to a range in orbital phase of $2\theta$ as
shown, the $y$ displacement returns to its starting point, requiring no
linear motion term be removed.  Anticipating our discussion of the
edge-on situation, we refer to this as the ``Worst'' astrometric case in
the sense it produces the largest unmodeled displacements.

For this case
\begin{equation}
2\varepsilon_y = a_p-l =\left \{ \begin{array}{r@{,\quad}l}
		     2a_p & P < \Delta t_m\\
		     a_p[1-\cos\frac{\pi\Delta t_m}{P}] & P \geq \Delta t_m
	      \end{array}\right .
\end{equation}
where $\Delta t_m$ is the mission lifetime, nominally 5 years.  For long
periods this becomes to first order
\begin{equation}
\varepsilon_y \sim a_p\left(\frac{\pi \Delta t_m}{2 P}\right )^2 
	    =  \left (\frac{\pi\Delta t_m}{2} \right )^2 \frac{m_s}{a^2}.
\end{equation}

\begin{figure}
\plotone{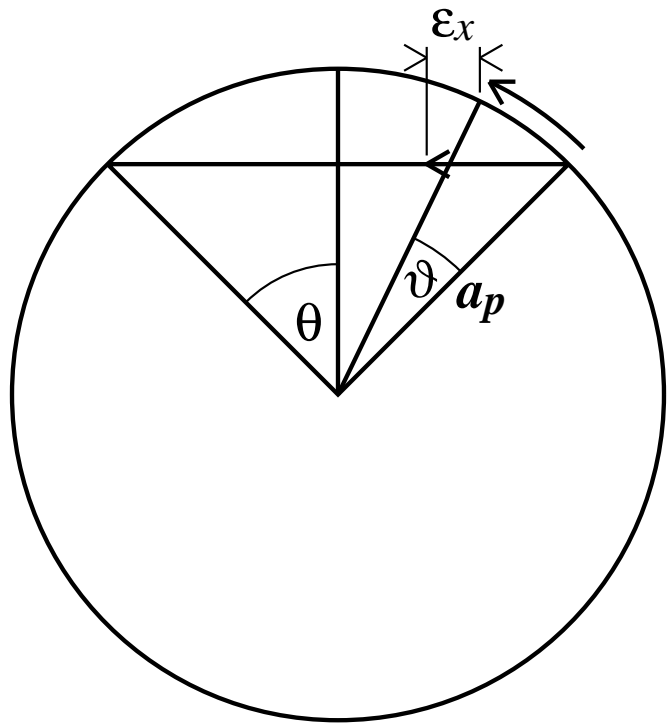}
\caption{\label{Circle2} The astrometric ``Best'' case.  Uniform linear
motion differs from projected uniform circular motion by third order terms.
The linear motion has been exaggerated for illustration.}
\end{figure}

The $x$ component involves a net translation.  The difference between
this and a linear translation is the noise term.  The geometry is shown
in Fig.\ \ref{Circle2} where the error term is the difference between an
average linear motion and the projected uniform angular motion.  Again,
$2 \theta$ will be traversed in the mission lifetime while $\vartheta$
corresponds to a specific epoch:
\begin{equation}
2\theta = \frac{2\pi\Delta t_m}{P},\quad \vartheta=\frac{2\pi(t-t_0)}{P}.
\end{equation}

From the construction we see that the apparent difference from uniform
motion is
\begin{equation}
\Delta x(\vartheta) =
a_p \left\{
\sin\theta - \sin(\theta-\vartheta) -\frac{\sin\theta}{\theta}\vartheta 
\right\}
\end{equation}
Extrema occur at
\begin{equation}
(\vartheta-\theta)_{max} = \pm\cos^{-1}\left[\frac{\sin\theta}{\theta}\right].
\end{equation}
Substituting these we find
\begin{equation}
\varepsilon_x =\left \{ \begin{array}{r@{,\quad}l}
		a_p & P < \Delta t_m\\
		a_p\left\{\frac{\sin\theta}{\theta}\cos^{-1}
		\left(\frac{\sin\theta}{\theta}\right)
		-\sin\left[\cos^{-1}\left(\frac{\sin\theta}{\theta}
		\right)\right]\right\}
		& P \geq \Delta t_m.
	      \end{array}\right .
\end{equation}
At large periods the latter becomes
\begin{equation}
\varepsilon_x \sim \frac{a_p\theta^3}{9\sqrt{3}} 
              = \left(\frac{\pi\Delta
              t_m}{9\sqrt{3}}\right)^3[m_p(1+q)]^{1/2}\frac{m_s}{a^{7/2}}.
\end{equation}

Note that asymptotically the ratio of Best (\ie\ $x$ component) to Worst
cases becomes $\varepsilon_x/\varepsilon_y = 4\pi\Delta t_m / 9\sqrt{3}P
\sim 0.8 \Delta t_m$.  The numerical factor differs by less than 3\% from 
the corresponding ratio in the radial velocity calculation, above.

\subsubsection{Edge-on Case}

As for radial velocities, the orbital phase, the angular distance of the
primary from the line of nodes, determines the magnitude of the
astrometric noise.  As can be seen in Figs.\ \ref{Circle1} and
\ref{Circle2}, inclining the system about the $x$ axis projects away the
Worst of the astrometric noise, leaving the Best component intact.
Exactly the opposite happens when projecting the system about the $y$
axis which removes the smaller, Best, component, leaving the Worst
undiminished.  Since asymptotically the Best to Worst ratio has
essentially the same numerical factor as for radial velocity detection
and because we assume the preparatory period, $\Delta t_p$ is the same
as the nominal mission, $\Delta t_m$, that is 5 years, the dependence of
the astrometric noise on orbital phase closely tracks that of the
extremes in the radial velocity sensitivity, {\it including the area
required for, say, 90\% detection}.  For clarity in the figures below,
we shade those limits only for radial velocity detection.

\subsubsection{Finite Inclination}

While in the edge-on case the estimation of astrometric noise and its
dependence on orbital phase closely parallels that of radial velocity
sensitivity, that is not true for inclination effects.  When the line of
nodes corresponds to the $x$ axis and the largest component of the
astrometric noise has been projected away, small deviations from an
exact edge-on orientation can easily restore that component to the point
that it dominates the error contribution.  For example, in the long
period case, when the ratio $\varepsilon_x/\varepsilon_y = 0.1$, that
is, when the period is 40.3\,yr (for a mission of 5 years) or $a =
11.7$\,AU (in the $m_s = 0$ limit), if the line of nodes is along the
$x$ axis, the projected (Worst) component will still be larger than the
Best for $\cos i \geq 0.1$, which will occur in 90\% of the systems.

Combining the effects of random orbital phases and system orientations
in estimating radial velocity sensitivity has the effect of broadening
the 90\% detection region, that is, of moving it farther to the left in
Fig.\ \ref{RadialVel}.  In contrast, combining these two effects in
estimating astrometric noise pushes the left boundary of the 90\%
detection region toward the right, asymptotically toward the Worst
locus.

As a result a reasonable ``rule of thumb'' for estimating the fraction
of astrometric variables missed when using radial velocity prefiltering
would be to compare the least sensitive limit of radial velocity
detection against the largest possible component of astrometric noise
when considering periods well larger than the preparatory time interval
and the mission lifetime.

\vfill\eject
\section{THE FILTERING PROBLEM}

We consider two archetypical populations from which Grid members could
be choosen.  The first are G dwarfs like the Sun, but nominally at
60\,pc.  These objects would have $V = 8.6$\,mag and angular diameters
of order 155\,\muas.  The second are weak lined (\ie\ Pop II) K giants
with $\mbox{[Fe/H]} \sim -1.0$ and $M_V \sim -1$\,mag.  Estimates of
SIM's target-to-target setting time corresponds roughly to the time
required to accumulate $10^7$ photons on a $V = 12$\,mag object, which
we take as an approximate faint limit for a Grid object.  This would
place an $M_V = -1$\,mag object at about 4\,kpc (with no extinction).
These objects would have angular diameters of order 20\,\muas.  (We
consider only the ``high galactic latitude problem'' for the K giants,
\ie\ no extinction, and comment on the problems of low latitude Grid
objects, and the effects of lower luminosity objects at high latitude in
the discussion).

The angular diameters are important.  A simple estimate is that starspot
induced positional noise could be kept below 2\,\muas\ if photometric
variability is kept below $\Delta V = 0.05$ and 0.40\,mag, respectively,
for these two populations which, while not trivial, is easily achieved
in practice \citep[\cf][]{FekelHenry}.

\subsection{Observational Screening}

\subsubsection{Radial Velocities}

For radial velocity screening, defining the required measurement
precision is critical.  While more precision results in better screening
for companions, this comes at a price in observing time, with the gKw's
potentially quite costly.  By now we have a substantial base of
experience in scrutinizing dG's.  \citet{Cumming_etal} summarize the
current practice for these objects.  We adopt a ($3\sigma$) requirement
of 30\,m\,s$^{-1}$ per observation for the G dwarfs.

The literature is less extensive regarding the K giants, not to mention
the older populations.  \citet{HatzesCochran} \citep[also
see][]{Frinketal} indicate that, while the gK photospheres seem to be
intrinsicly noisier than dwarfs, it is rare to encounter amplitudes of
100\,m\,s$^{-1}$ (and some of that detected velocity variability might
be due to multiplicity).  We adopt as realistic a ($3\sigma$) detection
limit of 100\,m\,s$^{-1}$.

\subsubsection{Direct Imaging}

In addition to velocity screening it is possible to detect multiplicity
in the wider systems through direct imaging.  For main sequence
primaries, there is an advantage in looking in the near infrared since
secondaries, presumably also on the main sequence, will be redder,
reducing the dynamic range involved.  Fortuitously, the practice of
adaptive optics is advancing faster at those wavelengths, offering
higher spatial resolution at $2\,\micron$.  This has been exploited for
example by \citet{Patience98} in detecting close companions in the
Hyades using the 10\,m Keck instruments.

From this we conclude that companions as faint as $\Delta K \sim 5.5$ at
separations of 1\arcsec\ or more should be detectable.  Judging by the
error estimates \citep{Patience98}, the sensitivity declines somewhat to
$\Delta K \sim 4.0$ at 0\,\farcs2.  From there we assume detectability
declines steadily to $\Delta K ~ 0.0$ at 0\,\farcs02, half the
diffraction limit of a 10 meter telescope at K.  In applying these
detection criteria, we linearily interpolate the above limits in
log-separation versus magnitude and assume direct detection is
impossible at separations less than 0\,\farcs02.

\subsubsection{SIM}

We also consider how SIM itself can provide one last pass at catching
problem systems.  This has been partially described in \citet{LP00} and
in detail in \citet{LP02}.  The idea stems from a suggestion by
\citet{Wielen96} that companions would induce wavelength dependent offsets
in position to the extent there were color differences between the two
components.  The situation is more complicated in an interferometer, the
extent to which the overlapping wavetrains interfere determines not only
the amplitude but also the sign of the offset, and this changes with
wavelength whether or not there are color differences.

This filter would be applied by carefully analyzing the initial
observations by SIM of each Grid member.  Catching additional problem
systems with the initial observations and and eliminating them from
routine Grid observations could allow the Grid to start with unnecessary
redundancy that could be pared down at the earliest stages of the
mission.

Since it is difficult to simply characterize this type of filtering, we
model it explicitly and defer discussion until its application in the
Grid simulations below.  At that point we will assume that each
observation collects $10^7$ photons (\ie\ the signal obtained in 2
minutes for a $V=12$ object, as mentioned above) and then evaluate
whether the companion will be detected.

\subsection{Parameters Space Sampled}

\begin{figure}
\epsscale{.7}
\plotone{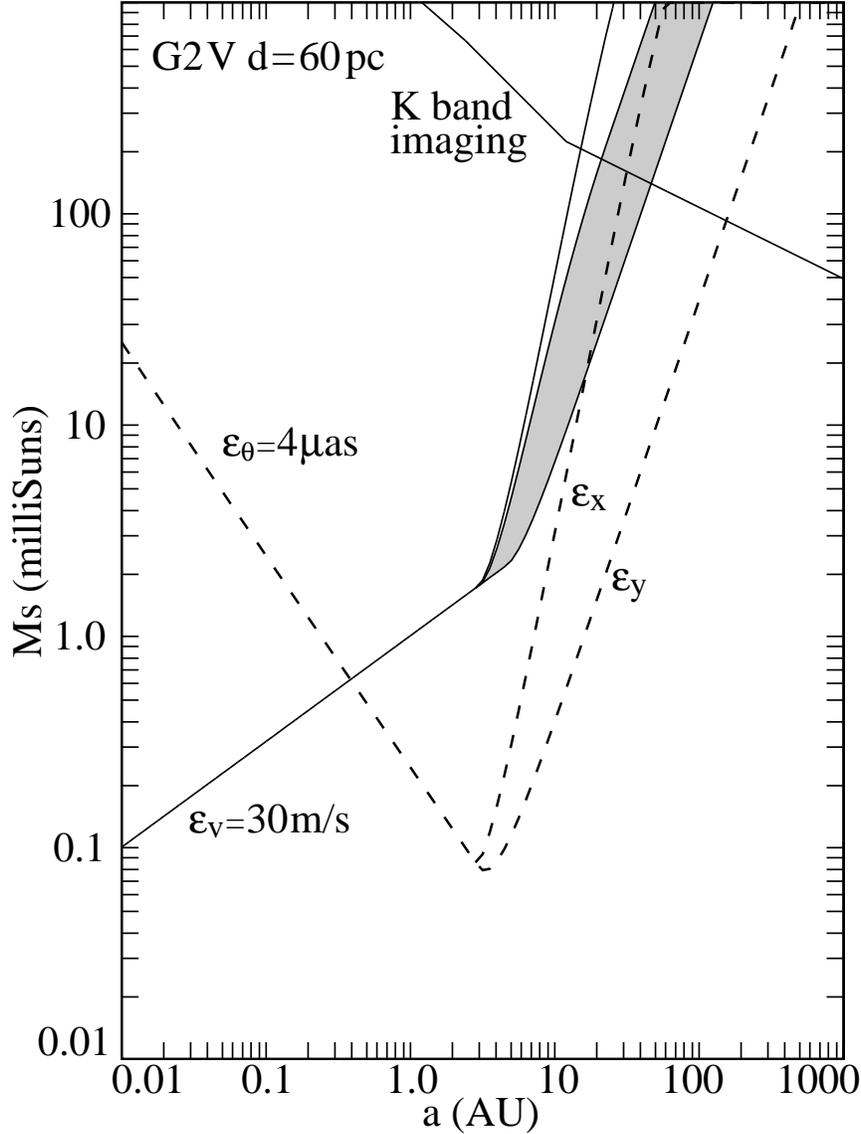}
\caption{This shows which nearby binaries with G dwarf primaries can be
detected with 30\,ms$^{-1}$\ radial velocity sensitivity ($\epsilon_V$)
versus which among the same systems would be astrometrically ``noisy''
at the $4\,\muas$ level ($\epsilon_{\theta}$).  The region where K-band
imaging could detect a companion is also shown.  The extreme orbital
phase effects are shown in both cases (at quadrature - rightmost - and
at conjunction - leftmost - for radial velocity detection, and along -
$\epsilon_x$ - and perpendicular to - $\epsilon_y$ - the line of nodes
for astrometric detection).  With regard to orbital phase, the region
that would provide 90\% velocity detection is shaded.  An essentially
identical region holds for astrometric detection and is not shown here
for clarity.  Systems below and to the right of the $\epsilon_V$ locus
would not be detected by radial velocity screening.  Systems above the
$\epsilon_{\theta}$ locus would produce unmodeled astrometric motions of
4\,\muas\ or greater.  \label{dGplot} }
\end{figure}

\begin{figure}[htb]
\epsscale{.8}
\plotone{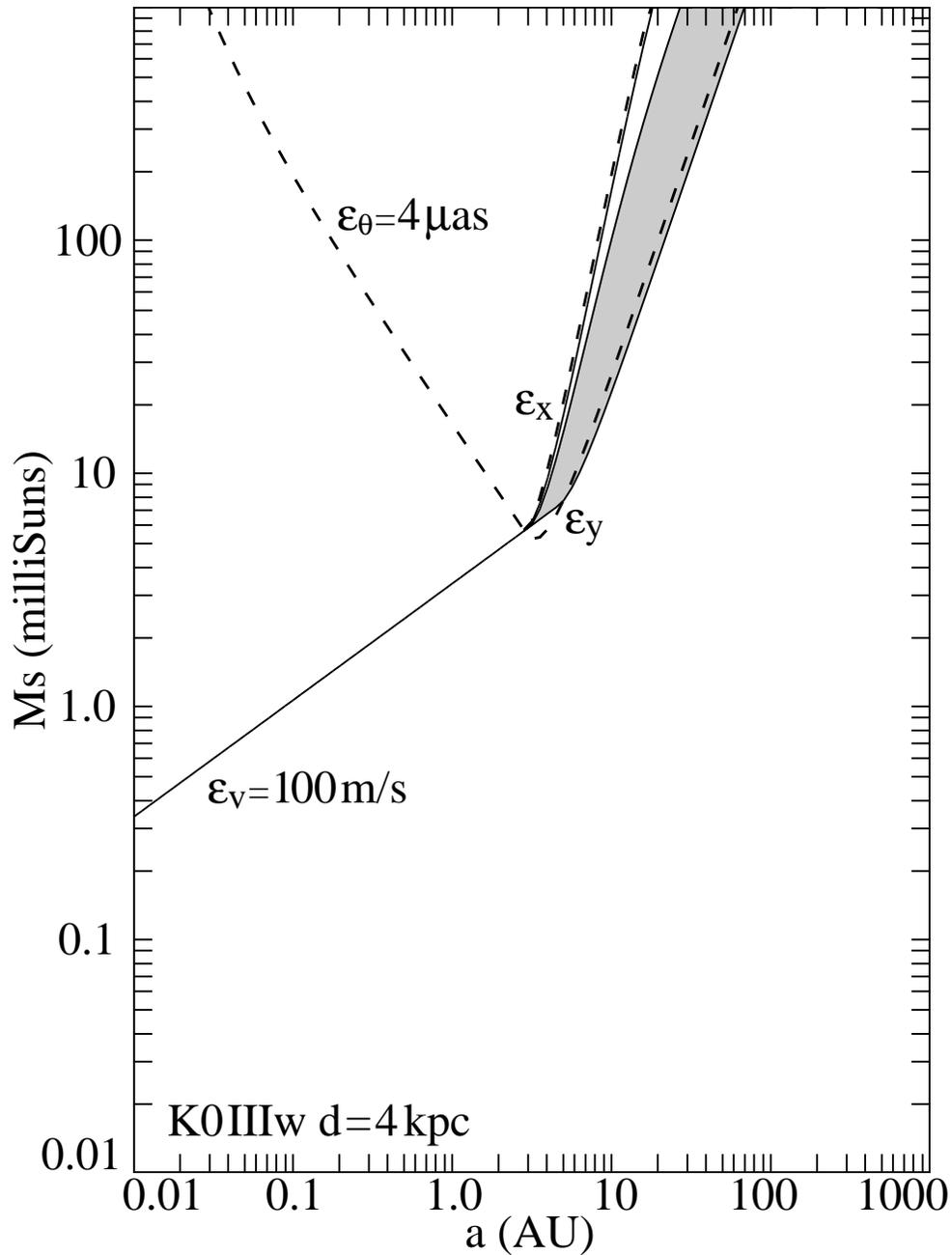}
\caption{The same as in Fig.\ \ref{dGplot}, only for 4\,kpc distant
weak-line K giants.  Here, K-band imaging no longer contributes.  A
substantially reduced requirement on the radial velocity screening
($\epsilon_V = 100$\,m/s) is applied.  Nevertheless, relatively few
systems displaying unmodelled astrometric motions with semiamplitudes in
excess of $4\,\muas$ (above the $\epsilon_{\theta}$ locus) would escape
radial velocity detection (below and to the right of the $\epsilon_V$
locus).  \label{gKwplot}}
\end{figure}

Figs.\ \ref{dGplot} and \ref{gKwplot} show the limits of radial velocity
binary detection compared to the regions occupied by astrometrically
``noisy'' systems for the two populations being considered here.
Objects above and to the left of the ``$\varepsilon_v= 30$\,ms$^{-1}$''
and ``$\varepsilon_v= 100$\,ms$^{-1}$'' loci will be detected as
binaries.  The effects of location in the orbit are indicated as
described above.  Also included in these figures are the ``Best''
($\varepsilon_x$) and ``Worst'' ($\varepsilon_y$) astrometric noise
cases that would be generated by these systems, labeled as
``$\varepsilon_\theta = 4\,\muas$''.  Systems with parameters above
these loci would exhibit 4\,\muas\ or more of unmodeled motion,
depending on inclination and orbital phase.  The region below and to the
right of the radial velocity loci, but above the astrometric loci
contains systems with potentially large unmodeled astrometric motions,
undetectable by radial velocity surveys of that precision.

Fig.\ \ref{dGplot} in addition shows the region of parameter space where
secondaries will be detectable by imaging methods.  In this figure we
assume the separation is the projected separation.  K-band imaging has no
application in the case of the K giants.  There, the brightest secondaries
are assumed to be main sequence stars of just below 1\,\Msun, that is G
stars that are even fainter compared to their primary at $2\,\mu m$ then
they are in the visible.  For G dwarf primaries K-band imaging is effective
at detecting the long period companions with mass ratios above about
$q=0.1$.

The differences between these two cases are striking.  The G dwarfs are
much more likely to be undetectable binaries, even with the help of
direct imaging.  However, the K giants are not completely immune to these
problems.  A note of caution in interpreting these figures: as we argue
below, secondary masses tend to be uniformily distributed (at least for
unevolved secondaries).  Plotting the log of the secondary mass results in
there being a large fraction of the graphed area where companions are
relatively uncommon.

However, a uniform distribution of mass ratios may not adequately describe
the situation at substellar masses.  A large number of spectroscopic
detections of companions to nearby solar type stars have been recently
reported with minimum masses in the neighborhood of a few milliSuns
\citep[\cf][]{Cumming_etal}.  If this closely reflects the true masses of
the companions, then there may well be a population of objects toward
the bottom of Fig.\ \ref{dGplot} that can contribute significant
astrometric noise.  Systems with distant gKw primaries would be immune
to the problems created by this population.

\vfill\eject
\section{DISTRIBUTION OF ORBITAL PARAMETERS}

\subsection{The Physical Parameters}

There is an extensive literature documenting efforts to deduce the
statistical distributions of the various parameters characterizing
binary systems.  Probably the most extensive recent such effort is that
of Duquennoy \& Mayor (hereafter DM) and colleagues \citep{DM, Mazeh,
Hogeveen}, whose results we adopt as our base set of distributions.

The three parameters of physical significance are the separation, the
eccentricity and the mass ratio, with the period a possible substitute
for the first through Kepler's third law.  For solar type primaries a
careful discussion led DM to conclude that the mass ratio is
approximately uniformly distributed over the range $1 \geq q
\geq 0$.  DM further concluded that the eccentricity distribution was
consistent with the probability density function $f(e) = 2e$ which is
expected from energy equipartition arguments \citep[\cf][]{Heggie}.

The remaining critical relation is the distribution of periods.  DM
adopted a broad Gaussian approximation to the log~P distribution they
found.  In this work we have elected to work with the distribution of
separations which is equivalent and which \citet{Heacox} has shown
yields the same functional form as for the periods, the only assumption
being that the masses are distributed independently of these parameters,
an assumption we adopt.  The variation in the probability density
function found by DM over the range of periods of interest is very
shallow and \citet{Heacox} has emphasized the small deviation of this
law from the simple power law, $f(P)\sim P^{-1}$ (or equivalently,
$f(a)\sim a^{-1}$).  Because we will later consider a substantially more
peaked distribution as an alternative, we deviate slightly from DM here
and adopt a pdf for separation that is uniform in $\log a$.

There is still considerable uncertainty in these distributions and the
possibility of significant variability depending on environment, age,
etc.  We therefore consider representative alternate density functions
for these three variables.  This will give us some sense of the
robustness of our results to the choice of those distributions.

Recently \citet{QL} (hereafter QL) have used the HIPPARCOS results to
constrain binary frequencies in the critical interval $0 \leq \log a
\leq2$, a range particularly difficult for both spectroscopic and direct
imaging at the distances of the typical HIPPARCOS objects.  They
concluded that the distributions adopted by DM for both separation and
mass ratio did not adequately describe the HIPPARCOS detections and
argued that the distribution for $\log a$ was much more sharply peaked
than proposed by DM (but peaked at about the same separations) and that
the mass ratio distribution increases substantially toward small mass
ratios.  We therefore adopt the following as the alternatives (to the
pdf's adopted by DM) for the semi-major axis and the mass ratio:

\begin{eqnarray}
f(\log a) = 
\frac{c}{\sqrt{2\pi}} \exp -\left( \frac{(\log a -1.5)^2}{2} \right); 
\quad -2 \leq \log a \leq 3\\
f(q) = 2(1-q); \quad 0 \leq q \leq 1.
\end{eqnarray}
In practice, the lower separation limit is never quite reached since we
limit the inner radius to $4 R_{\hbox{star}}$, which even in the case of
the Sun is about 0.02\,AU.

There are well known observational biases that make it difficult to
extract the probability density function for eccentricity.  We argue
that the data presented by DM are no stronger than ``consistent'' with
the functional form of the eccentricity density function they adopted.
On the other hand the binaries in well studied clusters like the Hyades
\citep[\cf][]{PS,Patience98} support a uniform distribution of
eccentricities (although neither set of authors made that argument).  We
shall therefore adopt a uniform density function for eccentricity as the
alternative pdf.

In general we assume that these three parameters, as well as the others
described below, are statistically independent.  We know that this
cannot be true in the limits of very small and very large separations.
In both cases there must be fewer large eccentricity cases than
otherwise.  For small separations, large eccentricities would take the
secondary into the primary, while for large separations, large
eccentricity orbits would be more easily tidally disrupted.  We ignore
the latter because we will be considering only relatively modest maximum
separations.

For small separations a combination of the details of the processes
leading to the birth of a binary and the subsequent effects of tides
lead to a substantial deficit of high eccentricity systems.  To allow
for these effects we limit the range of eccentricities for short period
G-dwarf systems by requiring that the periastron separation be no less
$4 R_{\hbox{star}}$.  Fortunately, the results of the simulations
described below are unaffected by the details of this perscription,
since essentially all such systems are detected according to the
criteria we have described.

The correlation of eccentricity and separation at short periods is more
complicated in the case of K giants, which we discuss next.

\subsection{Weak Line K Giants}

The above apply explicitly to disk F and G dwarfs, by far the most
extensively studied group in terms of binary parameters.  The situation
for gKw stars is poorly documented by comparison.  However, we believe
that we can, with care, extend the results described above to these
objects.  The critical fact on which the assumption hinges is the
discovery by Latham and coworkers \citep[][Latham 2000, private
communication, Latham, \etal\ 2002]{Carney94} that the binary frequency
in the F and G subdwarfs is indistinguishable from that of disk G
dwarfs.  That in stark contrast to the long held belief that Pop\,II
objects had about half the number of binaries compared to Pop\,I.
\citet{Latham2002} also provide evidence that the mass ratio
distributions for the two populations are similiar.

By far the most critical parameter in comparing binary populations is
their frequency.  Finding that the two populations have the same binary
frequency, it is not a large leap to argue that the other pdf's are
probably quite similar as well (save, perhaps, for more complete tidal
stripping, which does not affect us).  We adopt that hypothesis and
assume that binaries are basically the same, whether formed 10\,Gyr or
100\,Myr ago.

However, we are not dealing with main sequence G primaries, but K
giants.  Evolution off the main sequence will certainly have a
significant effect if for no other reason than that the primaries will
have radii approaching 30\,\Rsun, $\log a = -0.85$.  Further, it is not
at all clear that the K giant that is now the primary, was the primary
of the original system of main sequence objects; we allow for the
possibility of more than one evolutionary channel for creating binaries
with a K giant primary.  We consider simple evolutionary effects first.

\subsubsection{Evolution of the primary}

The first channel we consider for creating binaries with a K giant
primary is that of simple first time evolution of the original primary
toward the giant branch.  Specifically we assume we are dealing with an
object that is 10\,Gyr old and nominally 1\,\Msun.  We also assume the
secondary is then at least slightly below that mass and is still on the
main sequence.

We do not deal here with the case of a system consisting of two giants
(\ie very small mass differences).  Such systems are not all that
uncommon, judging by their frequency in the Bright Star Catalogue
\citep{Hoffleit} for example.  However, they are easily recognized and
will be rejected from the Grid immediately.  Further, their apparently
high frequency is in part the result of the relative brightness of the
system compared to a single giant; the volume probed by a magnitude
limited catalog such as the BSC for such systems approaches a factor of
three larger than that for systems dominated by the primary.  Since such
systems require mass ratios quite close to unity, their true frequency
is small and ignoring them will have no effect on our conclusions.

The effect of simple evolution of the primary will be to disrupt the
binary in those systems with initial separations less than about twice
the final radius of the primary.  Such systems will in general not
produce a simple field K giant and we discard them when such parameters
are generated in the Monte Carlo simulations.  For slightly wider
systems there will be a range of separations where initially eccentric
systems will undergo partial or complete circularization.

The calculation we adopt is slightly more complicated, and we proceed as
follows.  First, with the distance for the adopted absolute visual
magnitude (dependent on the metal deficiency) for the K giant and the
angular diameter from \citet{VanBelle} appropriate to the color of a
K0\,III, we calculate the primary's radius.  Then, if initially
periastron is less than 4 stellar radii we alter the eccentricity and
separation, conserving angular momentum \citep{NelsonEggleton}, so that
periastron is $4 R_{\hbox{star}}$ or, if that is not possible, the orbit
is circular.  In the latter case we then check whether the primary is
within its Roche lobe \citep[][but beware, Eggleton's ``$q$'' is our
$q^{-1}$]{Eggleton}.  If not, the system is rejected.

The upper limit on separation is maintained at $\log a = 3$.

\subsubsection{K giant secondaries}

On review of the various modes of binary evolution, one other channel
appears to provide a significant number of binary systems with K giant
primaries, although here the secondary, a white dwarf, was the system's
original primary.

In order for the original secondary to have evolved to a metal deficient
K giant, we argue that it must have been nominally 1\,\Msun.  For the
sake of this calculation we assume that all metal poor stars are about
10\,Gyr old and are coeval.  Further, we assume that it is highly
unlikely that the original secondary was significantly less than
1\,\Msun\ and then received just the amount of mass from the evolving
primary to make it enough more massive than 1\,\Msun\ that it evolved
into the giant phase now.  The original secondary will be taken as
having been 1\,\Msun\ all along.

Generalizing this argument, we assume that if there were any significant
mass transfer as the primary evolved it would not produce a system that
contains a Pop\,II K giant now.  This implies that we are dealing with
systems with separations of 10\,AU or more.  While approximate, this
cutoff is uncertain by no more than a factor of two, which is tolerable
given the small contribution of this channel.

Given these assumptions, we know that for Pop\,I an isolated star will
evolve to a white dwarf if it is initially 8\,\Msun\ or less (the
alternative, a more massive star going supernova, is a negligible source
of K giant systems).  In binaries where the primary can be stripped of
its hydrogen envelope at an early stage the lower mass limit for forming
neutron stars may be considerably higher, perhaps up to 12\,\Msun.
However, the mass functions are sufficiently steep that the upper limit
to masses contributing white dwarf secondaries has almost no effect on
the results.  We take 8\,\Msun\ as the upper mass limit and adopt it for
Pop\,II objects as well.

Over the range $1\,\Msun\leq m_i \leq 8\,\Msun$\ the primary loses mass
becoming a white dwarf of mass $m_f$. We adopt a simple power law to
describe the outcome,
\begin{equation}
m_f = 0.6m_i^{0.42},
\end{equation}
which gives a Chandrasekar mass white dwarf for an 8\,\Msun\ primary and
a 0.6\,\Msun\ white dwarf for a 1\,\Msun\ primary, as derived for
typical field white dwarfs.  We further assume that the frequency of
primary masses follows the Salpeter mass function and the mass ratio is
uniformly distributed in the original, unevolved binary.

The ratio of the number of systems from this second channel is related
to the number in the first channel (K giant primary) through the mass
function.  For the second channel the number density function of
primaries with mass, $m_p \pm dm_p$ and secondaries with mass ratio,
$q \pm dq$, is given by
\begin{equation}
        dn = \phi(m_p,q) dm_p dq = k m_p^{-2.35} dm_p dq,
\end{equation}
where the constant $k$ is for normalization.  We convert this to a
distribution of the secondary mass and integrate over primary masses of
1 to 8\,\Msun\ to obtain:
\begin{equation}
        \Delta n_s = 0.42 k dm_s.
\end{equation}

In turn, when the K giant is the original primary, then from the same
mass function we find for the first channel (a 1\,\Msun\ K giant)
\begin{equation}
        \Delta n_p = k dm_p.
\end{equation}
Equating $dm_p$ from the normal channel to $dm_s$ for the white dwarf
channel we find
\begin{equation}
        dn(wd)/dn(normal)=0.42.
\end{equation}
(This ratio is sensitive to the mass ratio distribution.  Substituting
the linearly increasing function proposed by QL for example, the above
evaluates to 0.25).

We must account for discarding the close systems (those that undergo
mass transfer) from the white dwarf channel.  In terms of the nominal
distributions, the relevant one being the uniformly distributed $\log
a$, $-2\leq \log a \leq3$, only 40\% of the original systems survive ($a
\geq 10\,$AU) and the ratio above is reduced to 0.17 (or 0.10 for the
linear mass ratio pdf).  (Using the more sharply peaked distribution for
separations found by QL, the fraction of acceptable systems increases to
67\% and the ratio of the two channels increases to 0.28.
Interestingly, if both of the QL distributions, mass ratio and
separation, are adopted the overall ratio remains at 0.17).

\subsection{Geometric Parameters}

In addition to these three physical parameters there are several which
describe the position of the orbit in space and the phase of the orbit.
There is no evidence that these are distributed in any other way than
one would expect.  We therefore assume that the longitude of the
ascending node, $\Omega$, and the argument of periastron, $\omega$ are
uniformly distributed over (0,$2\pi$) and that orbital planes are
randomly oriented in space, leading to the well know pdf for
inclinations, $f(i) = \sin i$.  Finally, we assume each system is
randomly distributed in orbital phase.  That is, $t-t_P$ is uniformly
distributed over $(0,P)$.

\vfill\eject
\section{SIMULATIONS}

We have run a number of Monte Carlo simulations to estimate the extent
to which astrometrically unstable systems can be detected.  We varied
the assumed orbital parameter distributions as well as certain other
parameters to illustrate how sensitive our conclusions were to the
various uncertainties.  Each run involves simulation of $10^4$ binary
systems.  In every case, we ran more than one such simulation, mostly to
check that we would not inadvertently reproduce the results with some
``$3\sigma$'' values.  The results quoted and graphed are those for one
specific run and should be reliable at the 1\% level.

The parameters of each simulated binary system were assumed to represent
the state of the system at the nominal start of the mission.  To assess
whether the system would be detectable as a spectroscopic binary, we
simply compared the velocity semi-amplitude of the system to our
threshold detectivity (\ie\ 30\,ms$^{-1}$ for G dwarfs) if the period
was shorter than the nominal 5-year preparatory time.  For periods
longer than 5 years we used half the total velocity change that would
have taken place the previous 5 years.  No allowance was made for
aliasing (which can modulate detectivity significantly, but only over a
small area in parameter space).  Also, no attempt was made to model the
radial velocity detection process -- a semi-amplitude was either larger
than the detection threshold, or not.

K-band detection was determined taking the nominal apparent separation
and the K magnitude difference deduced from the masses and comparing
them to the criteria defined above.  Again, no effort was made to model
the detection process in a statistical sense.

Only in the case of evaluating whether SIM would detect the presence of
a companion did we make a detailed model of the detection process.  This
results in the occasional ``detection'' of a system that is unlikely to
be detected, and vice versa (we refer to this as ``leakage'').  The
effect is small and will be remarked on only briefly below.

To assess astrometric stability we also distinguished systems with
periods longer than or shorter than the nominal mission life, 5 years.
For shorter periods, we evaluated the nominal apparent ellipse, calling
the system ``unstable'' if the semimajor axis exceeded $4\,\muas$.  For
longer periods we took 20 samples, evenly spaced, of the $x$ and $y$
components of the system's position over a mission lifetime, removed a
linear term, and again looked for any residuals that exceeded $\pm
4\,\muas$.  We note that a semi-amplitude of $4\,\muas$ is approximately
$2.8\,\muas$, rms.

\subsection{The Circular Approximation}

\begin{figure}
\epsscale{.7}
\plotone{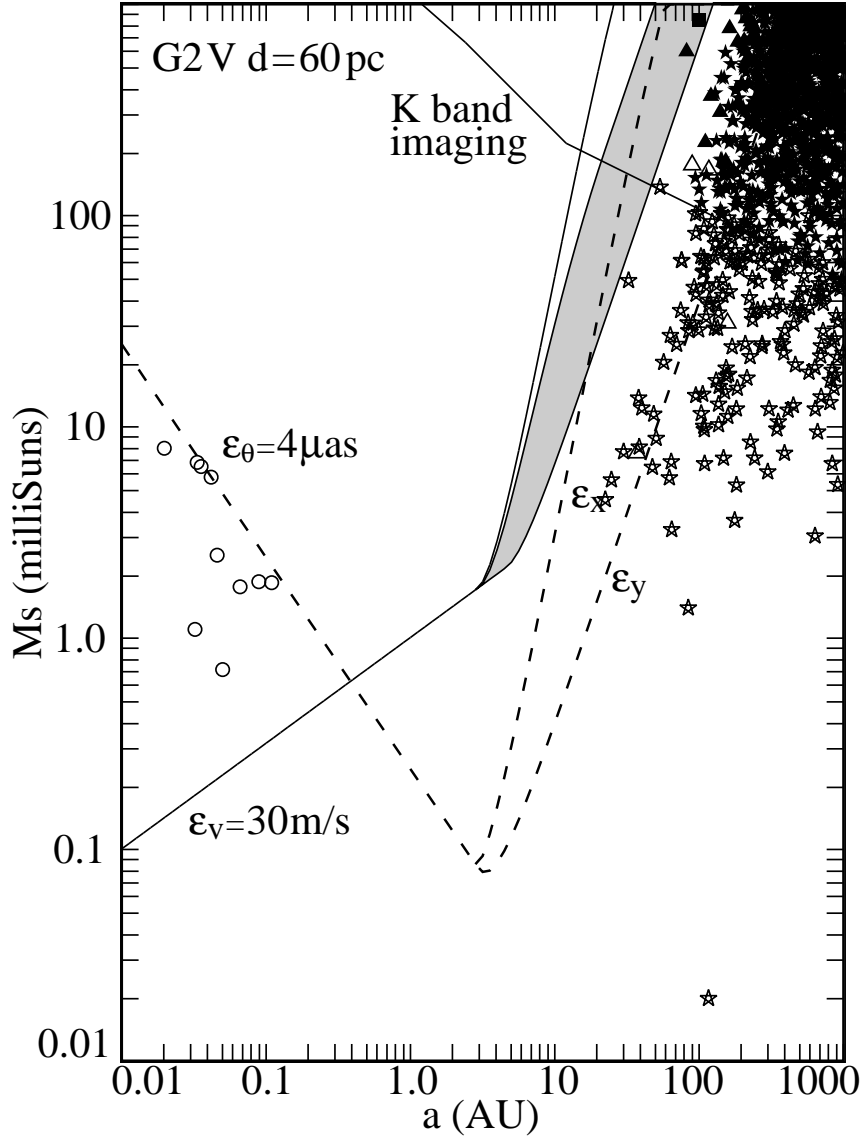}
\caption{\label{Fig:G2Vstable} The astrometrically ``stable'' systems
for $10^4$ simulated binaries with G2\,V primaries at 60\,pc, shown as a
function of semimajor axis and secondary mass.  The symbols indicate
whether the systems would have been detected as binaries and if so, how:
open circle ($\bigcirc$) for Rv only detectable, open box ($\square$)
for Rv + SIM only detectable, and open triangle ($\bigtriangleup$) for
SIM only detectable.  Filled versions of those symbols indicate the
systems would be detectable by K-band imaging, too.  A filled star
($\bigstar$) indicates detectable by K-band imaging only, while an open
star indicates a system undetectable by any of the techniques
considered.  The circular approximation does an excellent job
representing the regions of astrometric stability (below the
$\epsilon_{\theta}$ locus) and radial velocity detectability (above the
$\epsilon_V$ locus).}
\end{figure}

\begin{figure}[htb]
\epsscale{.8}
\plotone{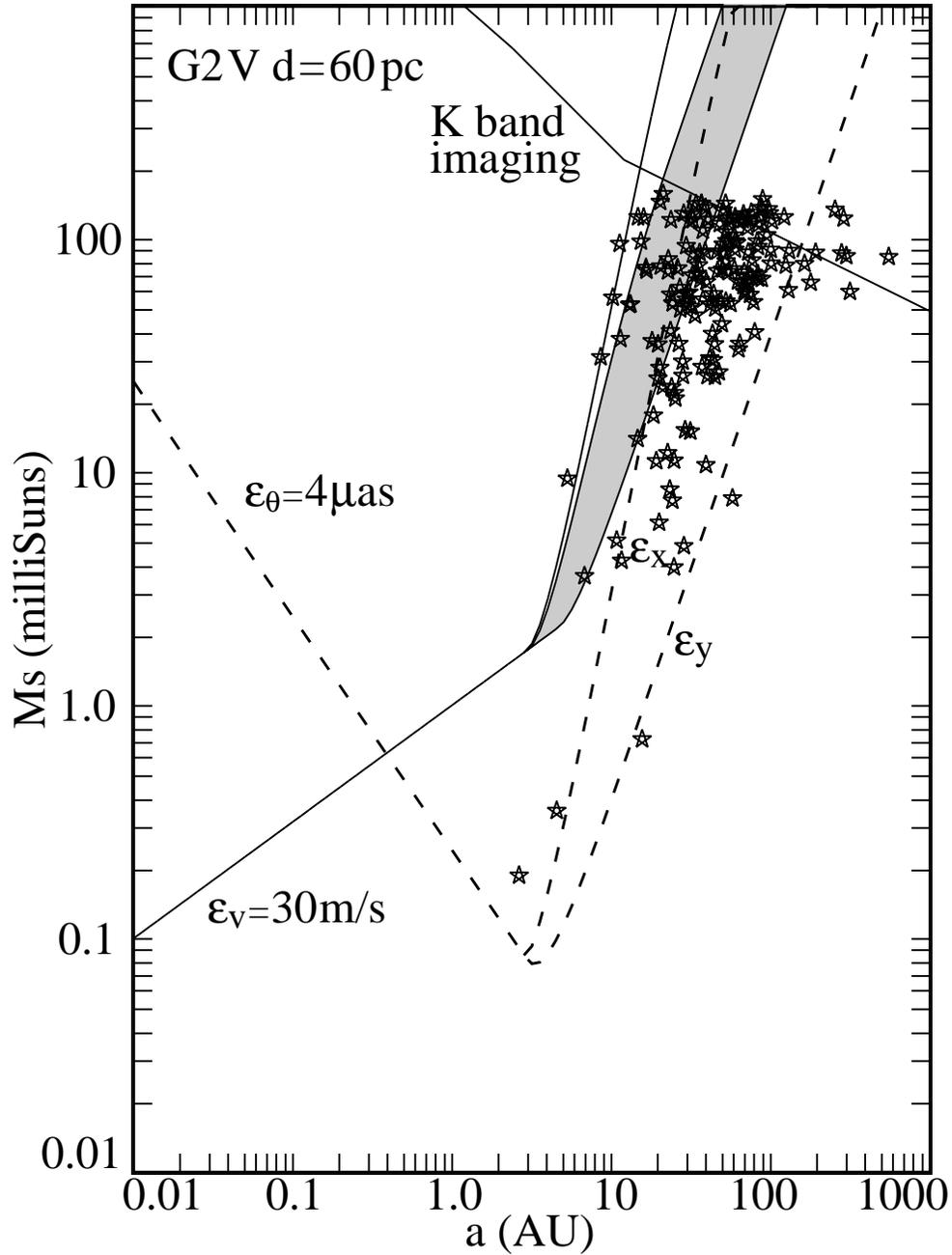}
\caption{\label{Fig:G2VunsUndet} As for Fig.\ \ref{Fig:G2Vstable} only
for the astrometrically unstable systems (above the $\epsilon_{\theta}$
locus) that are in principle not detectable by radial velocity screening
(to the left of the $\epsilon_V$ locus) or K-band imaging (above and to
the right of that locus).}
\end{figure}

We first compare the results from the simulations to the analytic
results from the circular orbit approximation.  Figures\,
\ref{Fig:G2Vstable} and \ref{Fig:G2VunsUndet} show the results for the
simulation for G dwarfs using the nominal distributions for astrometrically
stable systems and for those unstable systems that would not be detectable
using radial velocity screening and/or K-band imaging.  Although the high
density of plotted points somewhat obscures the boundaries in the former,
it is clear that the circular orbit approximation gives a good estimate of
which binary systems are detectable by radial velocity techniques and which
binary systems will display deleterious unmodelled motions.

In Fig.\ \ref{Fig:G2Vstable} we also indicate stable systems which would
in principle be detectable by K-band imaging and in Fig.\
\ref{Fig:G2VunsUndet} the sharp limits on lack of detectability of
unstable systems.  Here, the issue is the extent to which the semimajor
axis approximates the mean separation.  As can be seen, this too is an
excellent approximation both in the mean \citep[the mean separation
should be about $0.95a$, \cf][]{Leinert93}, and in the small dispersion
about that mean.

\begin{figure}
\plotone{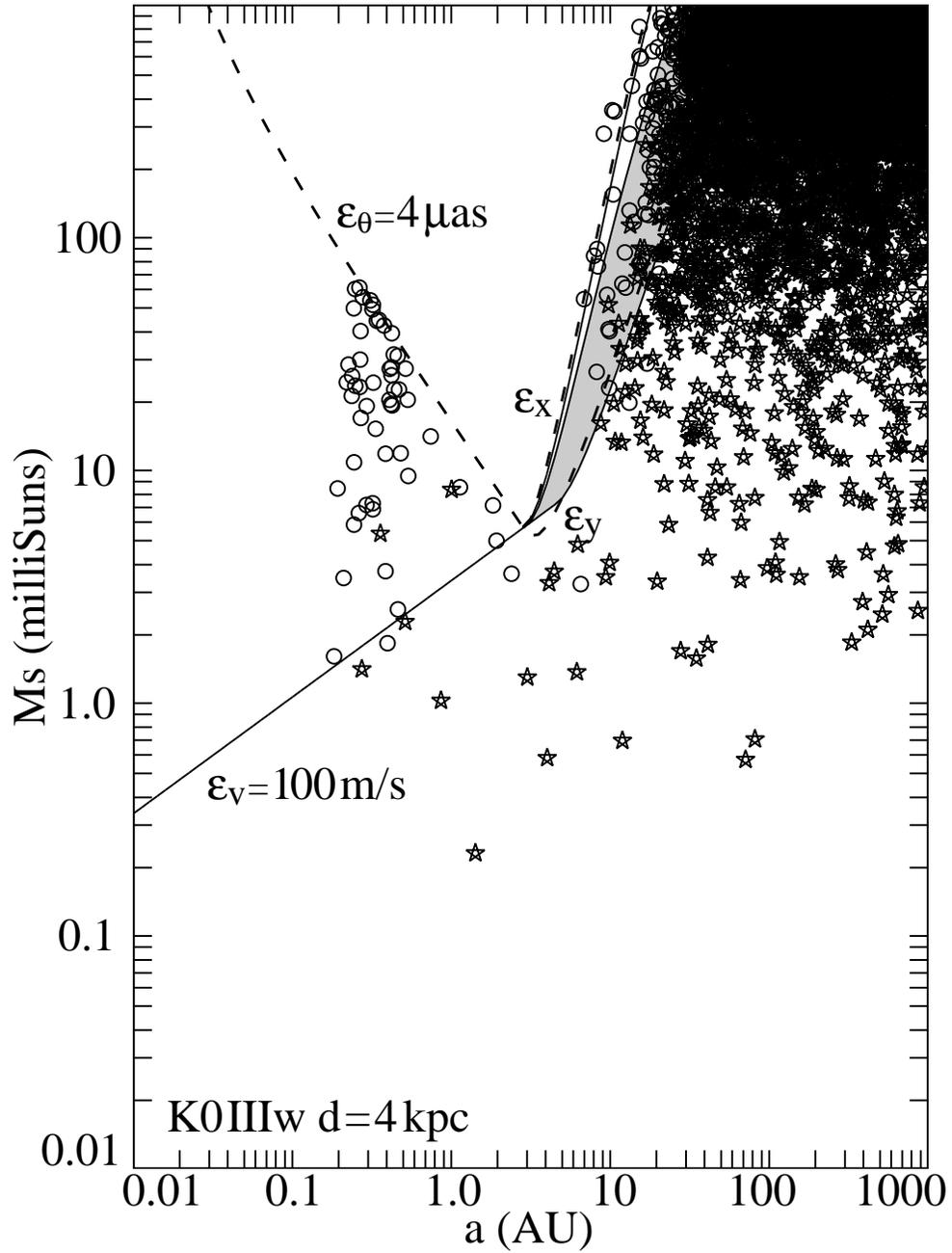}
\caption{\label{Fig:K03wstable} The weak-lined K giants that would be
astrometrically stable.  Symbols as in Fig.\ \ref{Fig:G2Vstable}.}
\end{figure}

\begin{figure}
\epsscale{.8}
\plotone{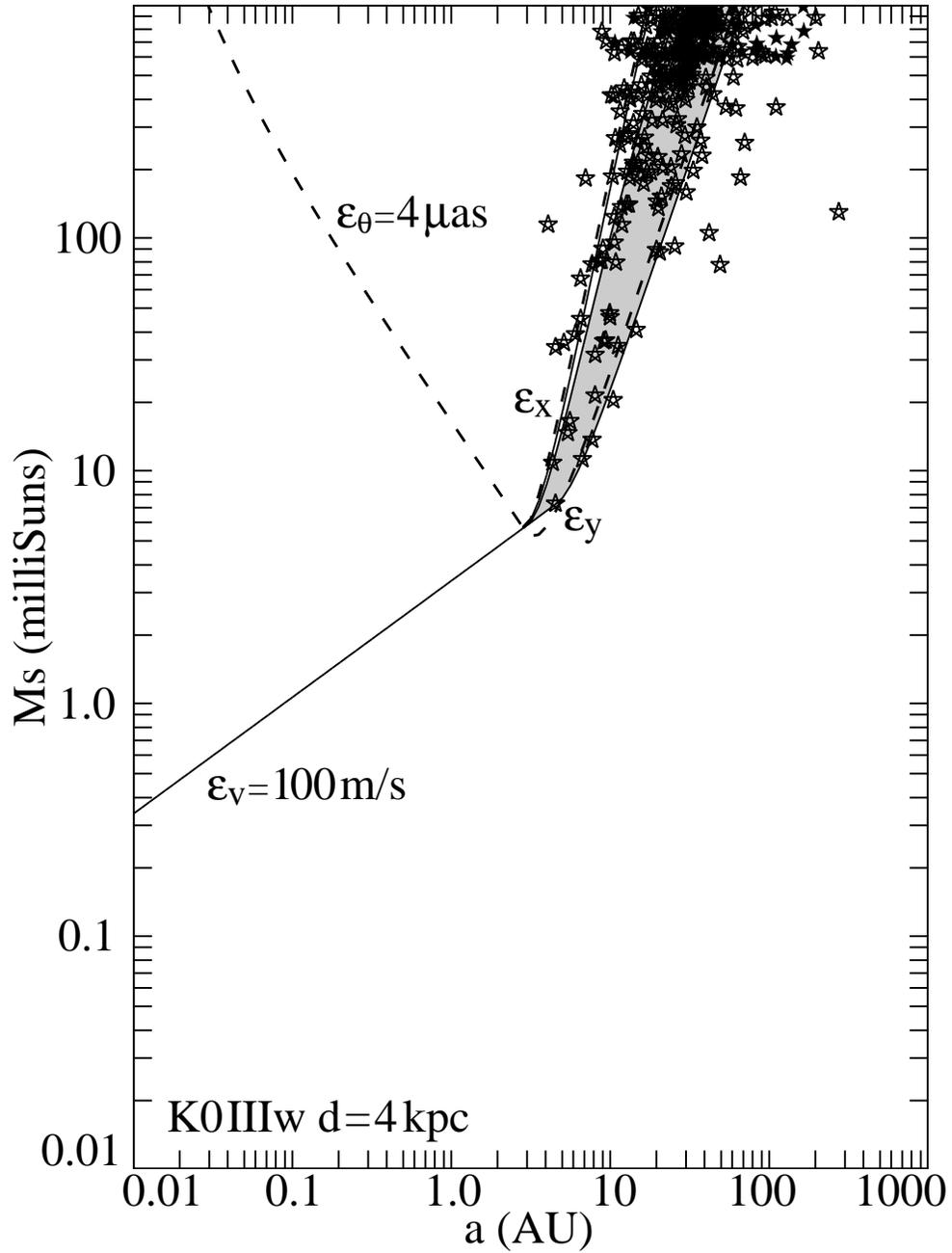}
\caption{\label{Fig:K03wunstable} As for Fig.\ \ref{Fig:K03wstable} only
for the astrometrically unstable (above the $\epsilon_{\theta}$ locus)
but radial velocity undetected (below the $\epsilon_V$ locus) systems.
Open stars indicate undetected systems with main sequence companions and
filled stars indicate such systems with white dwarf companions.}
\end{figure}

Figures \ref{Fig:K03wstable} and \ref{Fig:K03wunstable} show the
astrometrically stable and unstable, undetected systems, respectively,
for the Pop II K giants.  Again, the detectability and stability regions
are in close agreement with the simple circular orbit predictions.  As
remarked earlier, the fraction of unstable but undetected systems is
somewhat smaller for the K giant case.  The really notable difference is
the far larger fraction of stable systems in this latter case.

We next consider in more detail the simulation results for these two
populations. 

\subsection{The G Dwarfs}

The results from the individual simulations for nearby (60\,pc) disk G
dwarfs are given in Table\,\ref{Tab:G2V}.  Column (1) gives a reference
number (to identify entries in the Venn diagrams) and (2) a phrase
describing the subset.  The results from the simulations with various
parameter distributions come next, with ``nominal'' (3) basically the DM
(as slightly modified) choice, ``$f(e) = 1$'' (4) for a uniform
eccentricity distribution, ``$\log a$'' (5) for the fairly peaked
semimajor axis distribution of QL, and ``$q$'' (6) for a distribution of
mass ratios rising (linearly) to small ratios.  The remaining columns
are described below.

The various population subsets listed are generally ``exclusive''.  That
is, to get the number of K-band detections with no radial velocity
detections, but independent of being detectable by SIM, one adds the
results of lines 13 and 14.  The number of systems not detected,
independent of their suitability as Grid members is found by adding
lines 4 and 7, and so forth.

\begin{figure}
\plotone{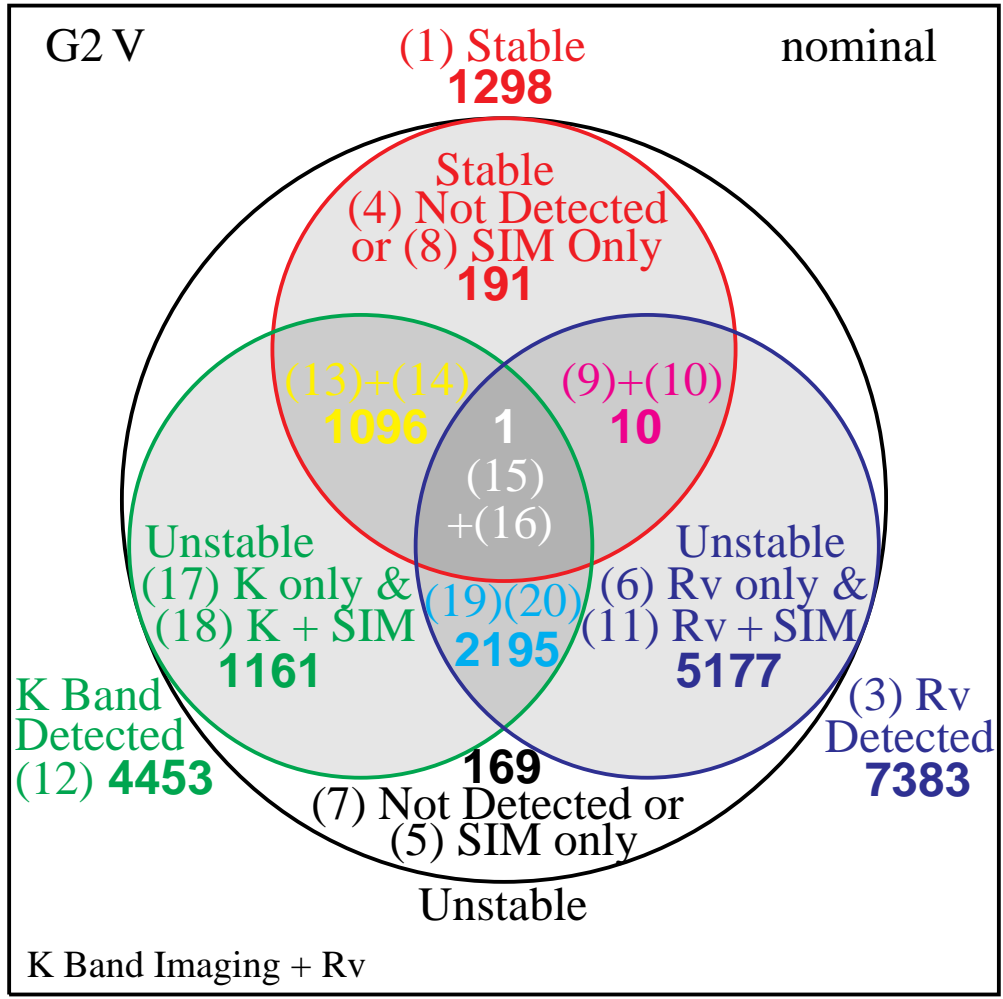}
\caption{\label{Fig:VGKR} The Venn diagram showing the detectability of
astrometrically stable and unstable systems with G2 dwarf primaries.
The effectiveness of K-band imaging in combination with radial velocity
screening in detecting companions is shown, independent of any
contribution SIM might make.  Numbers in parentheses index entries in
Table \ref{Tab:G2V}.}
\end{figure}

\begin{figure}[htb]
\plotone{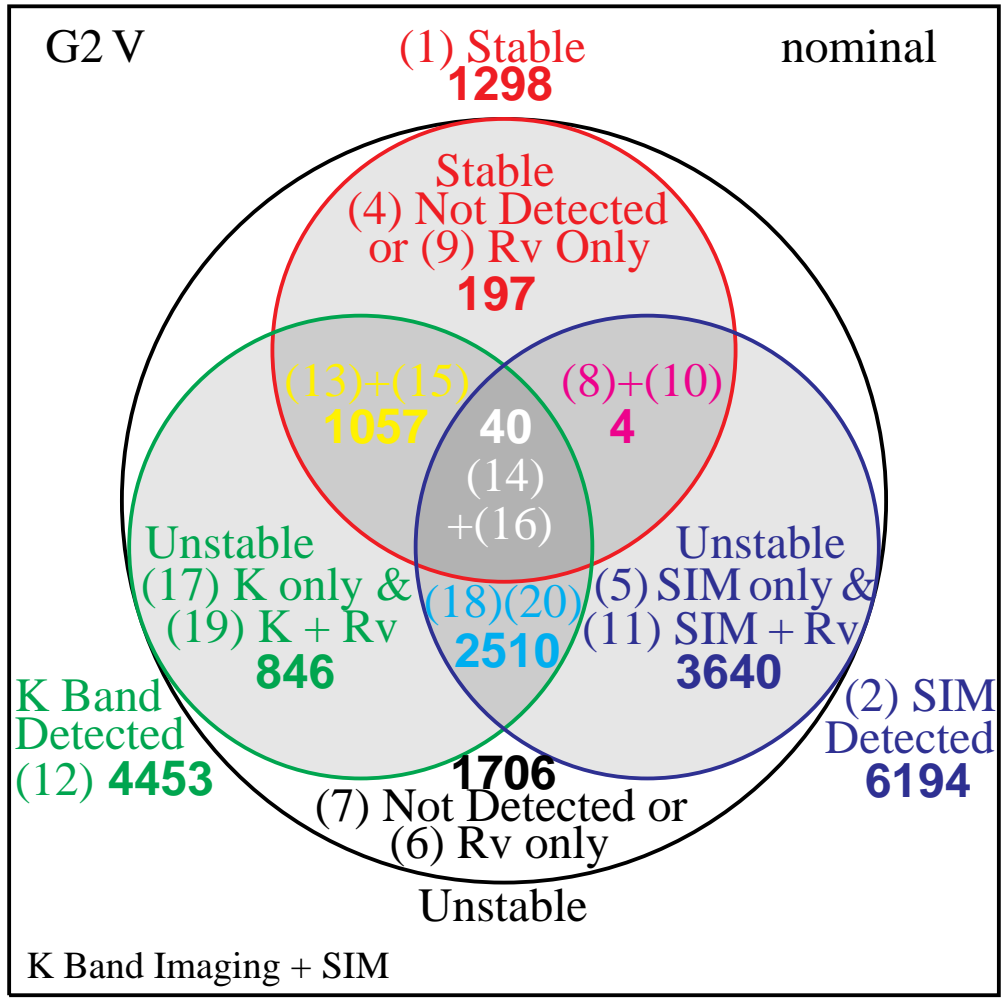}
\caption{\label{Fig:VGKS} The Venn diagram showing the effectiveness of
K-band imaging and an ultimate screening by SIM in detecting companions
in systems with G2 dwarf primaries, independent of any radial velocity
screening.}
\end{figure}

\begin{figure}[htb]
\plotone{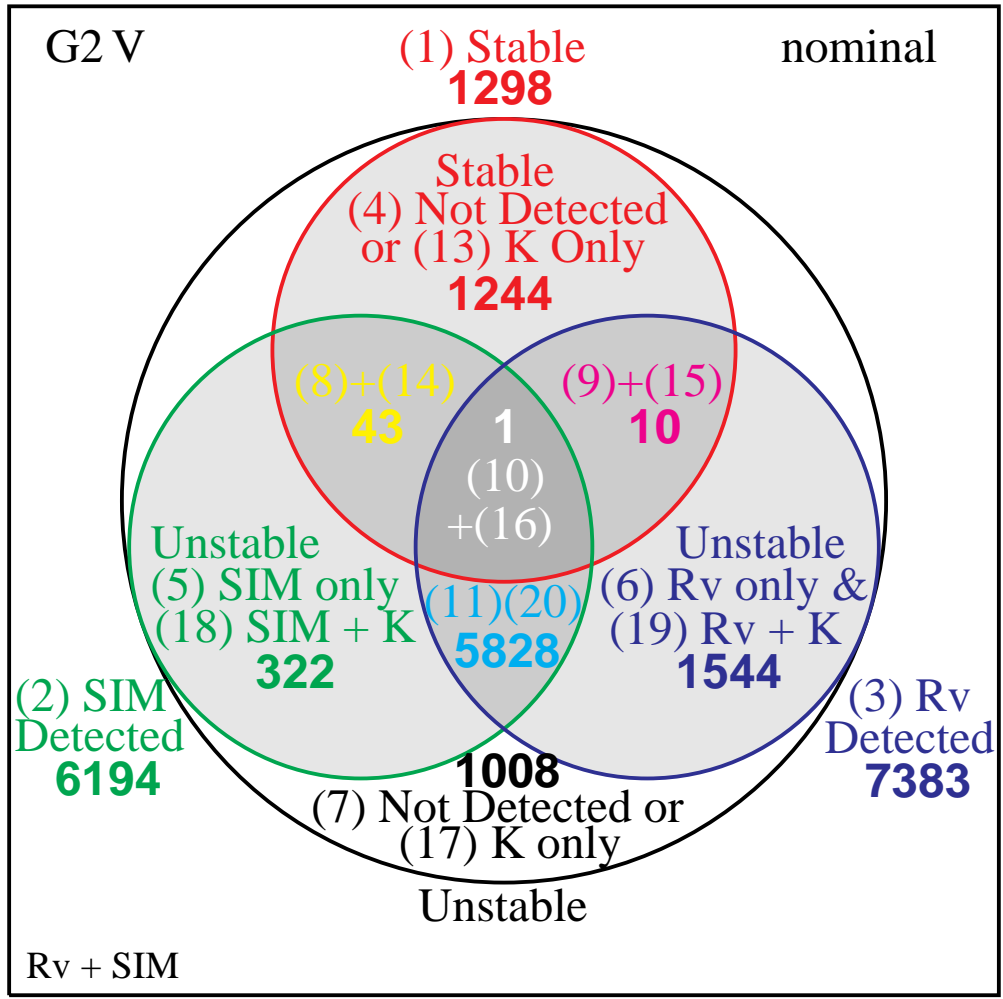}
\caption{\label{Fig:VGRS} The Venn diagram showing the effectiveness of
discerning radial velocity variations plus an ultimate screening by SIM
in detecting companions in systems with G2 dwarf primaries, independent
of any K-band imaging.}
\end{figure}

For the nominal distributions (column 3 of Table\,\ref{Tab:G2V}), these
results are shown in Venn diagrams, Figs.\,\ref{Fig:VGKR} --
\ref{Fig:VGRS}.  It is not possible to show the full overlap of
stability, K-band detection, RV detection, and SIM detection
simultaneously in such graphics, so we show the three possible
permutations, two detection techniques at a time.  We argue below that
SIM probably won't detect systems in unique parts of parameter space,
compared to radial velocity screening combined with K-band imaging, and
thus will concentrate on the combination presented in Fig.\
\ref{Fig:VGKR}.

The situation for the dG systems is that radial velocity screening
and K-band imaging combine to be remarkably efficient at detecting the
presence of a companion.  What is unexpected is the remarkably small
fraction of systems that are astrometrically stable.  This result is
robust over the range of parameter distributions we consider.  

Although we do not propose to get into the issue of optimizing the Grid
selection process, a naive approach would be to simply throw out all
systems detected as binary.  In the ``nominal'' example, that would
result in a yield of 191 stable systems and 169 unstable systems, a
devastating result.

Even if one could imagine some process of deciding which detected
binaries to retain from the radial velocity and K-band measurements that
was 100\% successful at removing the unstable systems, there would still
be a $\sim 13$\% contamination rate from undetected systems and a
discouragingly low $\sim 13$\% yield (1298 out of $10^4$).  The latter
would imply having to examine $1.5\,10^4$ potential targets to get a
yield of 2000 Grid members.

This, of course, assumes the initial population is 100\% binary, which
we discuss below.

\subsection{The weak-lined K giants}

The situation for the K giants is summarized in Table\,\ref{Tab:K03w},
which presents basically the same breakdown of the simulation results as
above.  K-band imaging makes no contribution here, at least for the
systems we consider, and those entries have been replace by a summary
of the contribution from white dwarfs.

\begin{figure}
\plotone{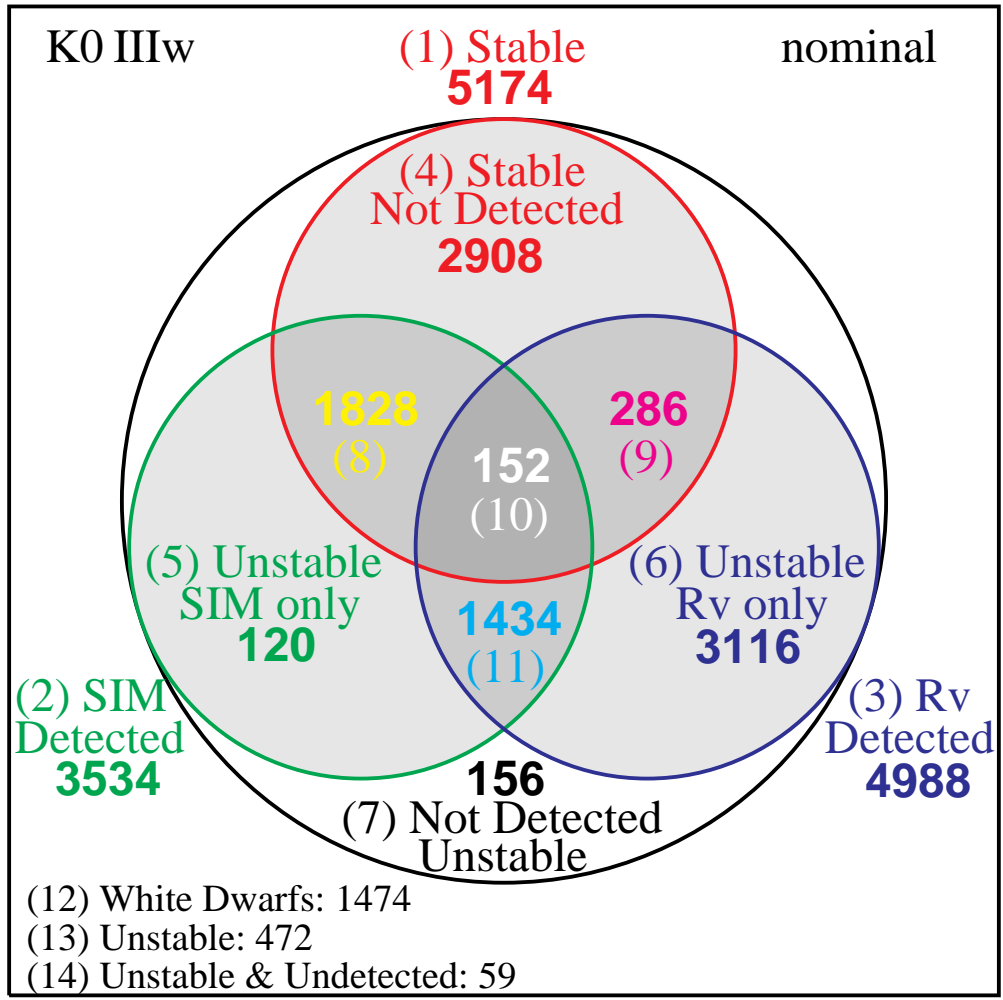}
\caption{\label{Fig:VK} The Venn diagram for systems with weak-lined K
giant primaries showing the dectability of a companion with SIM and
radial velocity screening and their overlap.  Numbers in parentheses
index entries in Table \ref{Tab:K03w}.  The contributions from systems
with white dwarf companions are summarized at the lower left.}
\end{figure}

Again the main conclusions are most easily drawn from the Venn diagram
shown in Fig.\,\ref{Fig:VK}.  Foremost is that there is the potential
for a significant yield, with about 52\% deemed astrometrically stable.
Even the issue of how to determine which are stable seems
straightforward.  If one ignores detectability by SIM and simply rejects
any system showing radial velocity variability, then 50\% of the systems
will be accepted, with only a 6\% contamination rate.  It is hard to see
how to improve on this much.

We note that the latter result is in complete agreement with the
estimate by \cite{Gould}, the addition of the white dwarf component does
not qualitatively affect the usefulness of the gKw population for the
SIM Grid.  On the other hand, nearly a third of the unstable but
undetected members of this population contain white dwarf secondaries.

Again, these conclusions are little affected by altering the various
parameter distributions, as seen in columns 3--6 of Table\,\ref{Tab:K03w}.
\cite{Gould} has noted that his estimates are most sensitive to the choice
of period distribution, which as noted is equivalent to the separation
distribution.  But we see little effect going from a pdf uniform in
$\log a$, as assumed in the ``nominal'' simulations, and the fairly
peaked distribution proposed by QL.  Consequently, we conclude that in
terms choosing Grid candidates, no one distribution stands out as
unusually poorly defined.

\subsection{The Role of SIM}

\begin{figure}
\plottwo{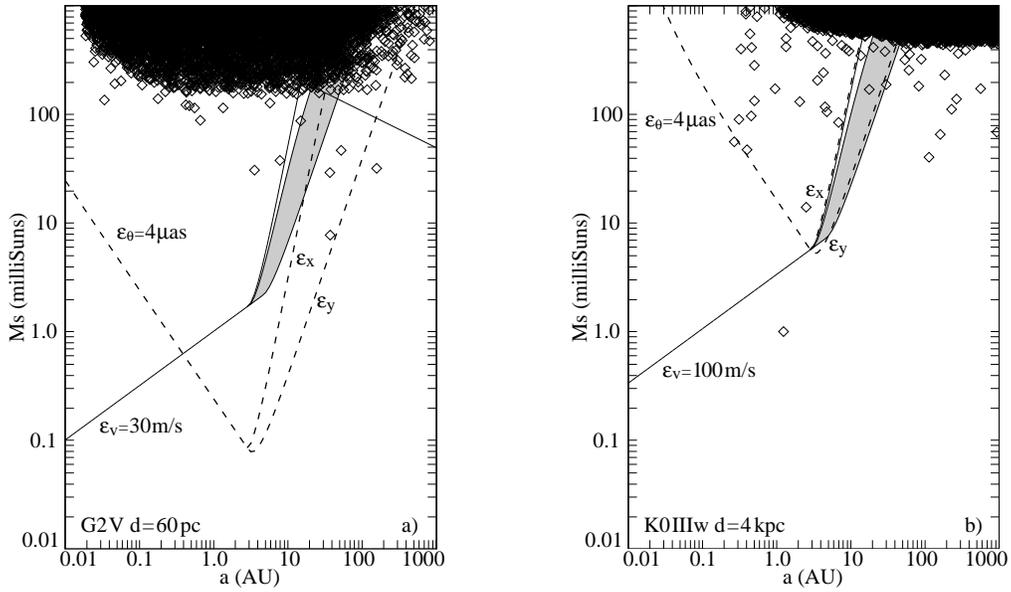}{figure14b.eps}
\caption{\label{fig:allSIM} Shown are all systems detected by the
simulation of the SIM measurement process (at the $2\sigma$ level) for
a) the G dwarfs and b) the K giants.  With the exception of some leakage
in the measurement process, essentially all unstable systems detected by
SIM are detectable using ground based methods.}
\end{figure}

As described in the Introduction, one of the goals of this investigation
was to discover whether SIM, itself, could contribute a useful final
screening for problem Grid members.  It appears that is unlikely.  In
Fig.\ \ref{fig:allSIM}a we show all dG systems detected by the SIM
simulation and in Fig.\ \ref{fig:allSIM}b those for the gKw systems.  In
the dG case, except for a little leakage in the simulation of the
detection process, essentially all SIM detections are capable of being
detected by either the radial velocity screening or K-band imaging.  For
the K giants, the coincidence of the radial velocity sensitivity limit
with the limits for the onset of astrometric instability imply that few
unstable systems would be detected by SIM without being detectable by Rv
screening.  These conclusions are borne out quantitatively in
Tables\,\ref{Tab:G2V} and
\ref{Tab:K03w}.

That said, we note that being ``detectable'' and actually being
``detected'' are quite different matters, and that a careful examination
of the SIM astrometric signal for an indication of multiplicity could
materially reduce contamination due to leakage in the radial velocity
screening process.

\subsection{Distances and Accelerations}

In assessing the weak-lined K giants we have taken as the ideal case a
substantially metal weak object, of order $[Fe/H] \sim -1$ in assuming $M_V
\sim -1$.  There are many more ``thick disk'' giants in this apparent
magnitude range than extreme Halo objects and it may on occasion be
required to rely on these not-so-distant objects in parts of the Grid.
To this end we have simulated the case for $M_V = 0.5$, where the
objects are still assumed to be $1\,\Msun$ (or, equivalently, this shows
the effects of 1.5\,mag of absorption for the higher luminosity
objects).  The results are shown in Table \ref{Tab:K03w}, column 7
(``$M_V$'').  These objects are at about 2\,kpc (unreddened) and there
is now a small range of separations where astrometric instability can
occur without significant probability of radial velocity detection.
Excluding all systems that could be detected by Rv screening, the yield
drops a little (4679) and the fraction of unstable systems in the
retained population rises to about 10\% (488).  Here it might be useful
to use the wide systems as seen by SIM and eliminate the closer ones to
reduce the contamination fraction.

However, there is a better way of proceeding.  \cite{Jacobs} has
suggested that at little loss in degrees of freedom it might be possible
in the post-mission, data processing phase to identify those systems
that show curvature and selectively add quadratic terms to the proper
motion solutions.  We show in Table \ref{Tab:K03w} the effect of
introducing quadratic terms in two of the simulations: the ``nominal''
case (column 8, ``2-nom'') and for the lower luminosity case just
described (column 9, ``$2-M_V$'').

The results are striking.  In the nominal case we see a substantial
(30\%) increase in actually ``stable'' systems, and an even more
striking 40\% gain for the lower luminosity ($M_V = 0.5$) population.
More remarkably, in the practical example of simply excluding all
systems detectable as radial velocity variables, we revert to the yields
found previously (50\% and 47\%, respectively) but with miniscule (0.3\%
and 0.8\%, respectively) contamination by unstable systems.  Again, this
is completely consistent with the findings of \cite{Gould}.  Since at
most 10\% of the systems will benefit from this adjustment, one can
expect little loss in the accuracy or numerical stability for the Grid
in the data reduction.

We wondered about the sensitivity of these results to the choice of the
stability requirement.  How badly would the yields degrade if for
example we required $2\,\muas$ amplitude for ``stability''?  The results
from several simulations showed that the yields were surprisingly
insensitive to the choice of that parameter.  We show in Table
\ref{Tab:K03w}, column 10 (``$22M_V$'') the results for the most extreme
case: lower luminosity K giants, but with quadratic terms in the proper
motion solutions.  The yield (everything not seen as variable in the
radial velocity screening) remains the same and the contamination level
rises from 0.8\% (36) to 1.1\%.  We can expect that many areas on the
sky will have Grid members stable to very high precisions, which may
open additional opportunities for SIM investigators.

In Table \ref{Tab:G2V}, column 7 (``2-nom'') we show the extent to which
introducing quadratic terms into the proper motion reductions reduce the
high fraction of undetected, astrometrically unstable systems with G
dwarf primaries.  As with the K giants the improvement is dramatic.  The
number of stable systems potentially available increases by more than a
factor of 2 although it remains a fairly low 31\%.  Further the
contamination rate (lines 5 + 7) is a remarkable 0.6\%.  Again, we do
not address exactly how to identify the stable systems, a critical step.

Finally, we wondered how strongly the selection process would degrade
with reduced precision in the velocity screening and the results were
again quite encouraging.  For the K giants, taking the simulation with
the nominal distributions, modelling only linear proper motion terms and
retaining a $4 \muas$\ stability criterion, we find that dropping the
($3\sigma$) velocity criterion to 200\,ms$^{-1}$ doubled the ``unstable
but undetected'' contamination to about 3\% and increased the yield by
10\% (to about 55\%).  As before, introducing quadratic terms in the
proper motions reduced the contamination by an order of magnitude.
However, this brings us into an area that we have not treated carefully,
the radial velocity detection process, and we think it is premature to
present details.  The radial velocity detection process needs cardful
modelling before we can provide a realistic estimate of the
contamination rate, particularly given the very small fractions
calculated here.  

\vfill\eject
\section{CONCLUSIONS}
We have considered the question of how to choose reliable objects for
SIM's Grid, and have looked at two groups of objects, nearby ($\sim
60$\,pc) disk population G dwarfs and kiloparsec distant thick
disk/halo, weak-lined K giants, which have been mentioned in this
context.  A knowledge of binary frequency and how the parameters of
binary systems are distributed is critical in answering this question.
We have argued that what is currently known for nearby G stars applys
generally.  

However, evolutionary effects must be allowed.  The K giants will have
modified the orbits of close companions, or have removed them.  Further,
the K giant primary need not have been the original primary of the
system, and we estimate the fraction that will have white dwarf
companions now, the result of evolution of the original primaries, to be
around 15\%.

With these results we have found that the weak-lined K giants at
distances of the order of 4\,kpc when screened for radial velocity
variations at the 100\,ms$^{-1}$ level, will provide a sample of objects
that will work quite acceptably for the SIM Grid.  For the standard Grid
data reduction, including only linear terms in the proper motions, one
expects a contamination level around 6--10\%, depending on the exact
distance of the population.  Allowing quadratic terms in the proper
motions where necessary, unmodeled motions can be kept below $2\,\muas$
for all but 1\% of the Grid (a result first obtained by
\cite{Gould}).

We also considered the plausibility of using nearby G dwarfs for this
purpose and argued that they are susceptible to two significant failure
modes if they were to be used.  First, considering the circular orbit
approximation we found that there is a sizeable range at low mass ratios
where companions would produce significant astrometric motions but not
radial velocity variations.  If the usual distributions found for more
massive companions holds over the whole range, this would provide no
significant channel for contamination.  However, the spate of recent
discoveries of Jupiter mass companions to nearby G dwarfs suggests
extrapolation of those results to low masses may not be wise.

Further, even if the assumed parameter distributions are as found for
the systems with stellar mass companions, we showed that if that
population consists of 100\% binaries, then the rejection rate from
radial velocity and K-band screening will reach at least 70\%, requiring
the screening if a large number of potential candidates.  The use of
quadratic terms in the proper motion reduction would again leave almost
no contamination in the resulting sample.  However, up to half the Grid
would require fitting these additional terms, raising questions about the
effects on the Grid's numerical stability.  Critical to these concerns
is the fraction of G dwarfs that are single, a parameter that is poorly
known and could well be zero.

We do note that there are limitations on the use of K giants as Grid
members in the Galactic plane, due to obscuration.  It might be that in
this reduced area, use of G dwarfs with a relatively high redundancy
factor would be an acceptable alternative.

\acknowledgements
We wish to acknowledge our colleagues who have made substantive comments
on this work, particularly, Steve Unwin, Richard Wade and Chris Jacobs.
Nancy Goddard contributed substantially in the early stages of this
project.  SPZ is supported as a fellow of the Royal Dutch Academy of
Science (KNAW).  Support from JPL contracts 961014 and 1218431 is
acknowledged and appreciated.

\vfill\eject

\vfill\eject
\begin{deluxetable}{rlrrrrr}
\tabletypesize{\scriptsize}
\tablecaption{Stability and Detectability of G2\,V Binary Systems.
\label{Tab:G2V}}
\tablehead{
\colhead{ (1)}   & 
\colhead{ (2)}   & 
\colhead{ (3)}   & 
\colhead{ (4)}   &  
\colhead{ (5)}   &
\colhead{ (6)}   & 
\colhead{ (7)}  \\
\colhead{ Index} &
\colhead{ Binary Status} &
\colhead{ nominal} &
\colhead{$f(e)=1$} &
\colhead{$\log a$} &
\colhead{ $q$ }  &
\colhead{ 2-nom}}
\startdata
 1 & Stable                           & 1298  &  1276  &   1374 &  1550 &  3152 \\
 2 & SIM Detected                     & 6194  &  6180  &   6202 &  4772 &  6194 \\
 3 & Rv Detected                      & 7383  &  7397  &   6364 &  7136 &  7383 \\
 4 & Stable: Not Detected             &  187  &   194  &    267 &   385 &   336 \\
 5 & Unstable: SIM Detected only      &    2  &     4  &      5 &     4 &     0 \\
 6 & Unstable: Rv Detected only       & 1539  &  1573  &    842 &  2628 &  1482 \\
 7 & Unstable: Undetected             &  167  &   181  &    317 &   354 &    18 \\
 8 & Stable: SIM only Detected        &    4  &     5  &      9 &     6 &     6 \\
 9 & Stable: Rv only Detected         &   10  &    11  &      0 &    18 &    67 \\
10 & Stable: SIM + Rv only Detected   &    0  &     0  &      1 &     0 &     6 \\
11 & Unstable: SIM + Rv only Detected & 3638  &  3613  &   1779 &  3038 &  3632 \\
12 & K Detected                       & 4453  &  4419  &   6780 &  3567 &  4453 \\
13 & Stable: K only Detected          & 1057  &  1022  &   1036 &  1093 &  1895 \\
14 & Stable: K + SIM only Detected    &   39  &    43  &     61 &    48 &   322 \\
15 & Stable: K + Rv only Detected     &    0  &     0  &      0 &     0 &     4 \\
16 & Stable: K + SIM + Rv Detected    &    1  &     1  &      0 &     0 &   516 \\
17 & Unstable: K only Detected        &  841  &   833  &   1321 &   740 &     3 \\
18 & Unstable: K + SIM only Detected  &  320  &   321  &    620 &   234 &    37 \\
19 & Unstable: K + Rv only Detected   &    5  &     6  &     15 &    10 &     1 \\
20 & Unstable: K + SIM + Rv Detected  & 2190  &  2193  &   3727 &  1442 &  1675 \\
\enddata

\tablecomments{Col.\ (1): Identifies which quantities are used towards the totals
shown in Figs.\ \ref{Fig:VGKR}--\ref{Fig:VGRS}.  Col.\ (2): Indicates the (exclusive)
logic used in these totals. Col.\ (3): A simulation based on the ``nominal''
probability density functions for the various binary parameters as defined in the
text. Col.\ (4): A simulation using the ``nominal'' pdf's for the binary parameters,
but substituting a uniform distribution for eccentricity. Col.\ (5): A simulation
using the ``nominal'' pdf's for the binary parameters, but substituting the somewhat
peaked distribution of QL for the semi-major axis.  Col.\ (6): A simulation replacing
the ``nominal'' uniform mass ratio (``q'') pdf with one that rises linearly toward
small ratios.  Col.\ (7): A simulation using the ``nominal'' pdf's, but with
quadratic terms removed from the apparent proper motions before characterizing the
motions as stable or unstable.}

\end{deluxetable}

\vfill\eject
\begin{deluxetable}{clrrrrrrrr}
\tablecaption{Stability and Detectability of K0\,IIIw Binary Systems. 
\label{Tab:K03w}}
\rotate
\tabletypesize{\scriptsize}
\tablehead{
\colhead{(1)} &  
\colhead{(2)} &  
\colhead{(3)} & 
\colhead{(4)}  & 
\colhead{(5)}  & 
\colhead{(6)}  & 
\colhead{(7)}  & 
\colhead{(8)}  & 
\colhead{(9)}  & 
\colhead{(10)}  \\
\colhead{Index} & 
\colhead{Binary Status} & 
\colhead{nominal} &
\colhead{$f(e)=1$}& 
\colhead{$\log a$} & 
\colhead{$q$} & 
\colhead{$M_V$} & 
\colhead{2-nom} &
\colhead{2-$M_V$}&
\colhead{22$M_V$}}
\startdata
 1 & Stable                           &  5174  &  4982  &   5220   & 5611  &  4395  &  6834 & 6124  &  5858 \\
 2 & SIM Detected                     &  3534  &  3374  &   3698   & 1734  &  3607  &  3534 & 3607  &  3607 \\
 3 & Rv Detected                      &  4988  &  5147  &   4928   & 4624  &  5321  &  4988 & 5321  &  5321 \\
 4 & Stable: Not Detected             &  2908  &  2789  &   2930   & 3987  &  2556  &  3053 & 2818  &  2806 \\
 5 & Unstable: SIM Detected Only      &   120  &   116  &    135   &   55  &   198  &     5 &    8  &    12 \\
 6 & Unstable: Rv Detected Only       &  3116  &  3379  &   2774   & 3582  &  3400  &  2296 & 2681  &  2853 \\
 7 & Unstable: Undetected             &   156  &   185  &    247   &  243  &   290  &    11 &   28  &    40 \\
 8 & Stable: SIM Only Detected        &  1828  &  1763  &   1760   & 1091  &  1635  &  1943 & 1825  &  1821 \\
 9 & Stable: Rv Only Detected         &   286  &   273  &    351   &  454  &   147  &  1106 &  866  &   694 \\
10 & Stable: SIM + Rv Only Detected   &   152  &   157  &    179   &   79  &    57  &   732 &  615  &   537 \\
11 & Unstable: SIM + Rv Only Detected &  1434  &  1338  &   1624   &  509  &  1717  &   854 & 1159  &  1237 \\
12 & White dwarfs                     &  1474  &  1411  &   1418   & 1425  &  1383  &  1474 & 1383  &  1383 \\
13 & White dwarfs: Unstable           &   472  &   442  &    542   &  433  &   519  &   115 &  124  &   161 \\
14 & WD: Unstable + Undetected        &    59  &    74  &     82   &   57  &   115  &     2 &    5  &     5 \\
\enddata

\tablecomments{Col.\ (1): Identifies the quantities shown in Fig.\ \ref{Fig:VK}.  Col.\ (2): Indicates the
(exclusive) logic used in these totals. Col.\ (3): A simulation based on the ``nominal'' probability density
functions for the various binary parameters as defined in the text. Col.\ (4): A simulation using the
``nominal'' pdf's for the binary parameters, but substituting a uniform distribution for eccentricity. Col.\
(5): A simulation using the ``nominal'' pdf's for the binary parameters, but substituting the somewhat peaked
distribution of QL for the semi-major axis.  Col.\ (6): A simulation replacing the ``nominal'' uniform pdf for
mass ratio (``q'') with one that rises linearly toward small ratios.  Col.\ (7): A simulation using the
``nominal'' pdf's but assuming $M_V = 0.5$ (d = 2\,kpc).  Col.\ (8): A simulation using the ``nominal'' pdf's
and $M_V = -1$, but with quadratic terms removed from the apparent proper motions before characterizing the
motions as stable or unstable.  Col.\ (9): A simulation combining the use of quadratic terms in the proper
motions and the lower luminosity assumption of Cols.\ (7) \& (8).  Col.\ (10): A simulation with the same
assumptions and distributions as Col.\ (9), but increasing the astrometric requirement to 2\,\muas.}

\end{deluxetable}

\clearpage

\end{document}